\documentclass[11pt]{article}
\usepackage{graphicx,caption2,subfigure}
\makeatletter
\newcommand{\singlespacing}{\let\CS=\@currsize\renewcommand{\baselinestretch}{1.0}\tiny\CS}
\newcommand{\doublespacing}{\let\CS=\@currsize\renewcommand{\baselinestretch}{1.5}\tiny\CS}

\begin{document}
\title {Rapidity Spectra of Heavy Baryons in Nuclear Collisions at Various Energies :
A Systematic Approach}
\author { Goutam
Sau$^1$\thanks{e-mail: sau$\_$goutam@yahoo.com},P.
Guptaroy$^2$\thanks{e-mail: gpradeepta@rediffmail.com},A. C. Das
Ghosh$^3$\thanks{e-mail: dasghosh@yahoo.co.in}$\&$ S.
Bhattacharyya$^4$\thanks{e-mail: bsubrata@www.isical.ac.in
(Communicating Author).}\\
{\small $^1$ Beramara RamChandrapur High School,}\\
 {\small South 24-Pgs, 743609(WB), India.}\\
 {\small $^2$ Department of Physics, Raghunathpur College,}\\
 {\small P.O.: Raghunathpur 723133,  Dist.: Purulia(WB), India.}\\
  {\small $^3$ Department of Microbiology,}\\
    {\small Surendranath College, Kolkata-700009, India.}\\
    {\small $^4$ Physics and Applied Mathematics Unit(PAMU),}\\
 {\small Indian Statistical Institute, Kolkata - 700108, India.}}
\date{}
\maketitle
\bigskip
\begin{abstract}
This study aims at understanding the nature of measured data on the
rapidity spectra of some heavy baryons [$\Lambda$, $\Lambda$bar,
$\Xi^-$ \& $\Xi$bar$^+$] produced in the nuclear collisions at some
modestly high energies. Furthermore, our objective is also to build
up a comprehensive and consistent methodology to analyze the data on
this specific observable which has a very important place in the
domain of High Energy Physics (HEP). On an overall basis, our target
here attains a moderate degree of success even for production of
such rare secondaries. In addition to this, the limitations of such
an approach have also been pointed out.
\end{abstract}
\newpage
\doublespacing
\section{Introduction}
Production of heavy baryons in both high-energy proton-proton and
nuclear collisions is relatively quite low, in so far as the
qualitative aspects pertaining to the number-density and
production-cross section are concerned. So, these heavy baryons are,
in general, not studied very frequently in a model-based manner,
though their production is physically no less significant than all
other secondaries. Rather they are of great importance due to their
high mass content and some other specific quantum numbers. Of them
some have `strangeness', for which they assume somewhat special
status. In this paper we would concentrate on the studies of
rapidity spectra alone on production of $\Lambda$, $\Lambda$bar,
$\Xi^-$ and $\Xi$bar$^+$ particle production in Pb+Pb and some other
nuclear collisions at various (high) energies.
\par For over the last few years we had tried to build up an
approach and an alternative methodology to study the rapidity
spectra of the produced secondaries (both light and heavy) and
succeeded to a considerable extent. And now we make new strides
vis-a-vis the studies of the rapidity spectra here on production of
some strange heavies in a few Heavy Ion Collisions. In the next
section (Section 2) we give the outline of the approach and the
methodology. The results are delivered in Figures and Tables in the
Section 3. And finally, we present the conclusions in the last
section (Section 4).
\section{The Phenomenological Setting : General Outlook and the Method}
Following Faessler\cite{Faessler1},
Peitzmann\cite{Peitzmann1}, Schmidt and Schukraft\cite{Schmidt1} and
finally Thom$\acute{e}$ et al\cite{Thome1}, we \cite{De1,De2} had
formulated in the past a final working expression for rapidity
distributions in proton-proton collisions at ISR (Intersecting
Storage Rings) ranges of energy-values by the following
three-parameter parametrization, viz,
\begin{equation}
\frac{1}{\sigma}\frac{d\sigma}{dy}=C_1(1+\exp\frac{y-y_0}{\Delta})^{-1}
\end{equation}
where $C_1$ is a normalization constant and $y_0$, $\Delta$ are two
parameters. The choice of the above form made by Thom$\acute{e}$ et
al\cite{Thome1} was intended to describe conveniently the central
plateau and the fall-off in the fragmentation region by means of the
parameters $y_0$ and $\Delta$ respectively. Besides, this was based
on the concept of both limiting fragmentation and the Feynman
Scaling hypothesis. For all five energies in PP collisions the value
of $\Delta$ was obtained to be $\sim$ 0.55 for pions\cite{De1} and
kaons\cite{De2}, $\sim$ 0.35 for protons/antiprotons\cite{De2}, and
$\sim$ 0.70 for $\Lambda$/$\Lambda$bar and $\Xi^-$/$\Xi$bar$^+$. And
these values of $\Delta$ are generally assumed to remain the same in
the ISR ranges of energy. Still, for very high energies, and for
direct fragmentation processes which are quite feasible in very high
energy heavy nucleus-nucleus collisions, such parameter values do
change somewhat prominently, though in most cases with marginal high
energies, we have treated them as nearly constant.
\par Now, the fits for the rapidity (pseudorapidity)
 spectra for non-pion secondaries produced in the PP reactions at various energies are phenomenologically
 obtained by De and Bhattacharyya\cite{De2} through the making of
 suitable choices of $C_1$ and $y_0$. It is observed that for most of the secondaries the values of $y_0$
 do not remain exactly constant and show up some degree of species-dependence .
  However, for lamdas ($\Lambda$, $\Lambda$bar) and cascades ($\Xi^-$, $\Xi$bar$^+$) it gradually increases with energies and the energy-dependence
 of $y_0$ is empirically proposed to be expressed by the following relationship\cite{De1} :
\begin{equation}
y_0=k\ln\sqrt{s_{NN}}+0.8
\end{equation}
 \begin{figure}
\centering
\includegraphics[width=2.5in]{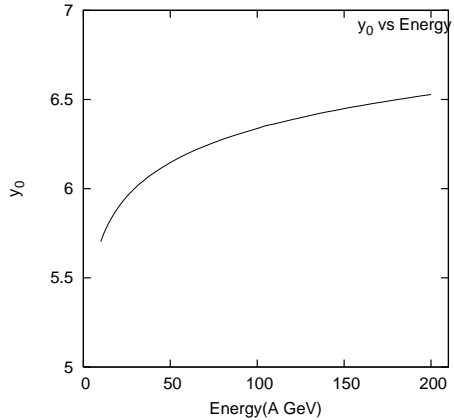}
\caption{Variation of $y_0$ in equation (2) with increasing energy.[Parameter values are shown in Table 1.]}
\end{figure}
\par The nature of energy-dependence of $y_0$ is shown in the adjoining figure
(Fig.1). Admittedly, as k is assumed to vary very slowly with c. m.
energy, the parameter $y_0$ is not exactly linearly correlated to
$\ln \sqrt{s_{NN}}$, especially in the relatively low energy region.
And this is clearly manifested in Fig.1. This variation with energy
in k-values is introduced in order to accommodate and describe the
symmetry in the plots on the rapidity spectra around mid-rapidity.
This is just phenomenologically observed by us, though we cannot
readily provide any physical justification for such perception
and/or observation. And the energy-dependence of $y_0$ is studied
here just for gaining insights in their nature and for purposes of
extrapolation to the various higher energies (in the centre of mass
frame, $\sqrt{s_{NN}}$) for several nucleon-nucleon, nucleon-nucleus
and nucleus-nucleus collisions. The specific energy (in the c.m.
system,
 $\sqrt{s_{NN}}$) for every nucleon-nucleus or nucleus-nucleus collision is first worked out by
 converting the laboratory energy value(s) in the required c.m. frame energy value(s).
 Thereafter the value of $y_0$ to be used for computations of inclusive cross-sections of nucleon-nucleon collisions
 at particular energies of interactions is extracted from Eq. (2) for corresponding obtained energies.
 This procedural step is followed for calculating the rapidity (pseudorapidity)-spectra for not only the pions
 produced in nucleon-nucleus and nucleus-nucleus collisions\cite{De1}. However, for the studies on the rapidity-spectra
 of the non-pion secondaries produced in the same reactions one does always neither have the opportunity to take recourse
 to such a systematic step, nor could they actually resort to this rigorous
 procedure, due to the lack of necessary and systematic data on
 them.
\par Our next step is to explore the nature of $f(y)$ which is envisaged to be given generally by a polynomial form noted
below :
\begin{equation}
f(y) = \alpha + \beta y + \gamma y^2,
\end{equation}
where $\alpha$, $\beta$ and $\gamma$ are the coefficients to be
chosen separately for each AB collisions (and also for AA collisions
when the projectile and the target are same). Besides, some other
points are to be made here. The suggested choice of form in
expression (3) is not altogether fortuitous. In fact, we got the
clue from one of the previous work by one of the authors
(SB)\cite{Bhattacharyya1} here pertaining to the studies on the
behavior of the EMC effect related to the lepto-nuclear collisions.
In the recent past Hwa et al\cite{Hwa1} also made use of this sort
of relation in a somewhat different context. Now let us revert to
our original discussion and to the final working formula for
$\frac{dN}{dy}$ in various AB (or AA) collisions given by the
following relation :
\begin{equation}
\frac{dN}{dy}|_{AB \rightarrow QX} = C_2(AB)^{\alpha + \beta y + \gamma y^2}\frac{dN}{dy}|_{PP \rightarrow QX}
 = C_3(AB)^{\beta y + \gamma y^2}(1 + \exp \frac{y-y_0}{\Delta})^{-1},
\end{equation}
where $C_2$ is the normalization constant and $C_3$=$C_2(AB)^\alpha$
is another constant as $\alpha$ is also a constant for a specific
collision at a specific energy. The parameter values for different
nucleus-nucleus collisions are given in the Tables (Table2 - Table
10).
\section{Results}
\subsection{A Few Pointed Steps}
The procedural steps for arriving at the results could be summed up
as follows :
\par (i) We assume that the inclusive
cross section (I.C.) of any particle in a nucleus-nucleus (AB)
collision can be obtained from the production of the same in
nucleon-nucleon collisions by multiplying the inclusive
cross-section (I.C.) by a product of the atomic numbers of each of
the colliding nuclei raised to a particular function, which is
initially unspecified\cite{Sau1}.
\par (ii) Secondly, we accept the property of factorization\cite{Sau1} of that particular function
which helps us to perform the integral over $p_T$ in a relatively
simpler manner.
\par (iii) Thirdly, we assume
a particular 3-parameter form for the pp cross section with the
parameters $C_1$, $y_0$ and $\Delta$.
\par (iv) Finally,
we accept the ansatz that the function f(y) can be modeled by a
quadratic function with the parameters $\alpha$, $\beta$ and
$\gamma$.
\subsection{Final Results Vis-a-vis the Measured Data-Sets}
The results are shown here by the graphical plots with the
accompanying tables for the parameter values. Here we draw the
rapidity-density of $\Lambda$, $\Lambda$bar, $\Xi^-$ and
$\Xi$bar$^+$ for symmetric Pb+Pb, C+C and Si+Si collisions at
several energies which have been appropriately labeled at the top
right corner. In this context some comments are in order. Though the
figures represents the case for production of $\Lambda$,
$\Lambda$bar, $\Xi^-$ and $\Xi$bar$^+$, we do not anticipate and/or
expect any strong charge-dependence of the results. Besides, the
solid curves in all cases-almost without any exception-demonstrate
our GSM-based results. Secondly, the data on rapidity-spectra for
some high-energy collisions are, at times, available for both
positive and negative y-values. This gives rise to a problem in our
method. It is evident here in this work that we are concerned with
only symmetric collisions wherein the colliding nuclei must be
identical. But in our expression (4) the coefficient $\beta$
multiplies a term which is proportional to y and so is not symmetric
under y$\rightarrow$(-y). In order to overcome this difficulty we
would introduce here $\beta$=0 for all the graphical plots. These
plots are represented by Fig.2 to Fig.14 for $\Lambda$,
$\Lambda$bar, $\Xi^-$ and $\Xi$bar$^+$ in Pb+Pb, C+C and Si+Si
collisions under different conditions of relatively low c.m. energy
values. The parameter values in this particular case are presented
in Tables (Table 2 - Table 10). The graphical plots shown in Fig.2 -
Fig.4 and Fig.7 (for $\beta$=0) are for production of $\Lambda$ and
$\Lambda$bar in Pb+Pb interactions at 20A GeV, 30A GeV, 40A GeV, 80A
GeV and 158A GeV respectively. And the plots depicted in Fig.5,
Fig.6, Fig.8 and Fig.9 are based on the data on production of
$\Lambda$ and $\Lambda$bar in Pb+Pb collision for the five different
centrality bins (for $\beta$=0) at 40A GeV and 158A GeV
respectively. The diagrams shown in Fig.10, Fig.11 and Fig.12
represent the production of $\Xi^-$ and $\Xi$bar$^+$ in Pb+Pb
collision at 20A GeV, 30A GeV, 40A GeV, 80A GeV and 158A GeV
respectively. And the diagrams in Fig.13 and Fig.14 represent the
production of $\Lambda$ and $\Lambda$bar in C+C and Si+Si collisions
at 158A GeV.
\section{Concluding Remarks}
The secondaries under study here are some heaviest of the heavy
secondaries on which measured data are quite sparse. Besides, the
errors/uncertainties [statistical plus systematic] in the
measurements of the data-sets are also quite considerable. In our
work we have reckoned with only the statistical errors alone and not
the systematic uncertainties, whereas the works by Alt et al and
Anticic et al have attached great importance to the systematic
uncertainties. Such differences in preference would make the
comparison of the obtained final results a bit difficult. From the
phenomenological points of view, the works by Alt et al and Anticic
et al are not much different, both of whom applied the sum of two
Gaussians placed symmetrically around mid-rapidity. However, the
present work makes use of a form which is not purely Gaussian and
has a clear but complex mass-number dependence of the nuclei in the
rapidity spectra. Still, the agreements between the measured data
and our model-based results are modestly satisfactory for both
$\Lambda$, $\Lambda$bar, $\Xi^-$ and $\Xi$bar$^+$ produced in nearly
all nuclear collisions. In contrast to the observations made by both
Alt et al and Anticic et al, we fail to obtain any clear
rapidity-plateau at even somewhat higher energies, though the trends
of the data-points have been faithfully reproduced by our approach.
The neglect of the systematic uncertainties in our approach might
play a crucial role in explaining this observed difference between
the nature of the rapidity spectra plotted by us and those plotted
by the sets of authors of Ref.\cite{Alt1} and Ref.\cite{Anticic1}.
Secondly, a comment is in order here with regard to the Fig.13a and
14a. We really fail to understand and account for the nature of
inversion of the rapidity-spectra in these two figures related to
the $\Lambda$-production in both C+C and Si+Si interactions at 158A
GeV. This remains a puzzle to us, as it does not conform to the
nature of all other plots, as depicted by the measured data or the
model-based ones. However, a common limitation of all the approaches
is that those cannot be applied for the non-symmetric collisions in
a straightforward manner. So, even the present approach cannot be
considered to be a fully generalized one. Despite this, it certainly
seems to provide a systematic approach to the studies on, at least,
all the symmetric collisions with a complex form of nuclear
mass-number dependence.
\begin{center}
\par{\textbf{Acknowledgement}}
\end{center}
\par
The authors express their thankful gratitude to the learned Referee
for his/her constructive comments and valuable suggestions in
improving the quality of an earlier version of the manuscript.
\newpage
\singlespacing

\newpage
{\singlespacing{
\begin{table}
\begin{center}
\begin{small}
\caption{Variation of $y_0$ with Energy.[Reference Fig. No.1]}
\begin{tabular}{|c|c|c|c|}\hline
 $Energy (A GeV)$& $\sqrt{s_{NN}}(GeV)$ & $Constant (k)$ & $y_0$ \\
 \hline
 $20$& $6.3$ & $ 2.76$ & $5.894$ \\
 \hline
 $30$& $7.6$ & $2.54$ & $6.006$\\
 \hline
 $40$& $8.7$ & $2.40$ & $6.085$\\
 \hline
 $80$& $12.3$ & $2.16$ & $6.276$\\
 \hline
 $158$& $17.3$ & $1.97$ & $6.463$\\
 \hline
\end{tabular}
\end{small}
\end{center}
%\end{table}

%\begin{table}
\begin{center}
\begin{small}
\caption{Values of different parameters for production of $\Lambda$
and $\Lambda$bar in Pb+Pb collisions at 20A GeV, 30A GeV, 40A
GeV(for $\beta$ = 0) for both +ve and -ve rapidities.[Reference Fig.
No.2-4]}
\begin{tabular}{|c|c|c|c|c|}\hline
$Energy(A GeV)$ & $Production$ & $C_3$ & $\gamma$ & $\frac{\chi^2}{ndf}$\\
 \hline
 $20$&$\Lambda$ & $14.169 \pm0.0734 $  & $-0.073 \pm0.0012$ & $9.075/05 $ \\
 \hline
 $20$& $\Lambda bar$ & $0.094\pm0.0015 $ & $-0.143 \pm0.0071 $& $0.159/03 $ \\
 \hline
 $30$&$\Lambda$ & $15.534 \pm0.0522 $  & $-0.049 \pm0.0004$ & $5.288/06 $ \\
 \hline
 $30$& $\Lambda bar$ & $0.216\pm0.0044 $ & $-0.104 \pm0.0048 $& $4.517/05$ \\
 \hline
 $40$&$\Lambda$ & $14.913 \pm0.0642 $  & $-0.035\pm0.0005$ & $7.540/05 $ \\
 \hline
 $40$&$\Lambda bar$ & $0.325\pm0.0049 $ & $-0.090 \pm0.0031 $& $5.169/06$ \\
 \hline
\end{tabular}
\end{small}
\end{center}
%\end{table}

%\begin{table}
\begin{center}
\begin{small}
\caption{Values of different parameters for production of $\Lambda$
in Pb+Pb collisions at 40A GeV(for $\beta$ = 0) for the 5 different
centrality bins C0-C4 for both +ve and -ve rapidities.[Reference
Fig. No.5]}
\begin{tabular}{|c|c|c|c|}\hline
 $Centrality$ & $C_3$ & $\gamma$ & $\frac{\chi^2}{ndf}$\\
 \hline
 $C0$ & $15.638 \pm0.0626 $  & $-0.039 \pm0.0004$ & $3.644/07 $ \\
 \hline
  $C1$ & $12.954 \pm0.0410 $ & $-0.039 \pm0.0007 $& $1.003/07 $ \\
 \hline
 $C2$ & $ 8.445\pm0.0407 $ & $-0.037 \pm0.0005 $& $1.922/03 $ \\
 \hline
  $C3$ & $5.409\pm0.0201 $ & $-0.026 \pm0.0005 $& $0.884/03 $ \\
 \hline
  $C4$ & $3.195 \pm0.0261 $  & $-0.026 \pm0.0009 $& $3.662/06 $\\
 \hline
\end{tabular}
\end{small}
\end{center}
%\end{table}

%\begin{table}
\begin{center}
\begin{small}
\caption{Values of different parameters for production of
$\Lambda$bar in Pb+Pb collisions at 40A GeV(for $\beta$ = 0) for the
5 different centrality bins C0-C4 for both +ve and -ve
rapidities.[Reference Fig. No.6]}
\begin{tabular}{|c|c|c|c|}\hline
 $Centrality$ & $C_3$ & $\gamma$ & $\frac{\chi^2}{ndf}$\\
 \hline
 $C0$ & $0.297 \pm0.0086 $  & $-0.076 \pm0.0066$ & $1.316/11$ \\
 \hline
  $C1$ & $0.234 \pm0.0074 $ & $-0.090 \pm0.0086 $& $2.429/11 $ \\
 \hline
 $C2$ & $0.184\pm0.0066 $ & $-0.108 \pm0.0124 $& $3.059/07 $ \\
 \hline
  $C3$ & $0.147\pm0.0029 $ & $-0.069 \pm0.0052 $& $1.871/09 $ \\
 \hline
  $C4$ & $0.094 \pm0.0024 $  & $-0.057 \pm0.0048 $& $0.307/04 $\\
 \hline
\end{tabular}
\end{small}
\end{center}
%\end{table}

%\begin{table}
\begin{center}
\begin{small}
\caption{Values of different parameters for production of $\Lambda$
and $\Lambda$bar in Pb+Pb collisions at 80A GeV, 158A GeV(for
$\beta$ = 0) for both +ve and -ve rapidities.[Reference Fig. No.7]}
\begin{tabular}{|c|c|c|c|c|}\hline
$Energy(A GeV)$& $Production$ & $C_3$ & $\gamma$ & $\frac{\chi^2}{ndf}$\\
 \hline
$80$& $\Lambda$ & $13.841 \pm0.0254 $  & $-0.013\pm0.0001$ & $0.572/03 $ \\
 \hline
 $80$ & $\Lambda bar$ & $0.771\pm0.0127 $ & $-0.025 \pm0.0031 $& $0.946/05$ \\
 \hline
 $158$&$\Lambda$ & $9.473 \pm0.0317 $  & $-0.006\pm0.0003$ & $1.829/04 $ \\
 \hline
 $158$&$\Lambda bar$ & $1.252\pm0.0069 $ & $-0.049 \pm0.0008 $& $7.249/05$ \\
 \hline
\end{tabular}
\end{small}
\end{center}
\end{table}

\begin{table}
\begin{center}
\begin{small}
\caption{Values of different parameters for production of $\Lambda$
in Pb+Pb collisions at 158A GeV(for $\beta$ = 0) for the 5 different
centrality bins C0-C4 for both +ve and -ve rapidities.[Reference
Fig. No.8]}
\begin{tabular}{|c|c|c|c|}\hline
 $Centrality$ & $C_3$ & $\gamma$ & $\frac{\chi^2}{ndf}$\\
 \hline
 $C0$ & $13.053 \pm0.0598 $  & $-0.010 \pm0.0006$ & $0.417/09 $ \\
 \hline
  $C1$ & $10.779\pm0.0318 $ & $-0.010 \pm0.0014 $& $0.039/03 $ \\
 \hline
 $C2$ & $ 8.296\pm0.0549 $ & $-0.019 \pm0.0005 $& $7.119/04 $ \\
 \hline
  $C3$ & $4.760\pm0.0085 $ & $-0.012 \pm0.0003 $& $0.119/03 $ \\
 \hline
  $C4$ & $3.269 \pm0.0216 $  & $-0.021 \pm0.0004 $& $0.288/02 $\\
 \hline
\end{tabular}
\end{small}
\end{center}
%\end{table}

%\begin{table}
\begin{center}
\begin{small}
\caption{Values of different parameters for production of
$\Lambda$bar in Pb+Pb collisions at 158A GeV(for $\beta$ = 0) for
the 5 different centrality bins C0-C4 for both +ve and -ve
rapidities.[Reference Fig. No.9]}
\begin{tabular}{|c|c|c|c|}\hline
 $Centrality$ & $C_3$ & $\gamma$ & $\frac{\chi^2}{ndf}$\\
 \hline
 $C0$ & $1.482 \pm0.0420 $  & $-0.049 \pm0.0032$ & $1.912/11 $ \\
 \hline
  $C1$ & $0.969\pm0.0123 $ & $-0.012 \pm0.0016 $& $0.253/05 $ \\
 \hline
 $C2$ & $0.775\pm0.0071 $ & $-0.021 \pm0.0015 $& $0.252/08 $ \\
 \hline
  $C3$ & $0.563\pm0.0175 $ & $-0.064 \pm0.0021 $& $1.490/05 $ \\
 \hline
  $C4$ & $0.380 \pm0.0041 $  & $-0.065 \pm0.0049 $& $1.388/05 $\\
 \hline
\end{tabular}
\end{small}
\end{center}
%\end{table}

%\begin{table}
\begin{center}
\begin{small}
\caption{Values of different parameters for production of $\Xi^-$
and $\Xi$bar$^+$ in Pb+Pb collisions at 20A GeV, 30A GeV, 40A
GeV(for $\beta$ = 0) for both +ve and -ve rapidities.[Reference Fig.
No.10 \& 11]}
\begin{tabular}{|c|c|c|c|c|}\hline
$Energy(A GeV)$& $Production$ & $C_3$ & $\gamma$ & $\frac{\chi^2}{ndf}$\\
 \hline
 $20$ &$\Xi^-$ & $0.961 \pm0.0073 $  & $-0.089\pm0.0072$ & $1.808/05 $ \\
 \hline
 $20$ & $\Xi bar^+$ & $\times $ & $\times$& $\times$ \\
 \hline
$30$&$\Xi^-$ & $1.213 \pm0.0269 $  & $-0.074\pm0.0053$ & $3.338/09 $ \\
 \hline
 $30$& $\Xi bar^+$ & $0.308\pm0.0051 $ & $-0.090\pm0.0029$& $0.509/03$ \\
 \hline
$40$ &$\Xi^-$ & $1.217\pm0.0373 $  & $-0.049\pm0.0035$ & $0.922/05 $ \\
 \hline
 $40$& $\Xi bar^+$ & $0.392\pm0.0105 $ & $-0.136\pm0.0046$& $1.019/03$ \\
 \hline
\end{tabular}
\end{small}
\end{center}
%\end{table}

%\begin{table}
\begin{center}
\begin{small}
\caption{Values of different parameters for production of $\Xi^-$
and $\Xi$bar$^+$ in Pb+Pb collisions at 80A GeV, 158A GeV(for
$\beta$ = 0) for both +ve and -ve rapidities.[Reference Fig. No.12]}
\begin{tabular}{|c|c|c|c|c|}\hline
 $Energy(A GeV)$&$Production$ & $C_3$ & $\gamma$ & $\frac{\chi^2}{ndf}$\\
 \hline
$80$& $\Xi^-$ & $1.577\pm0.0335 $  & $-0.036\pm0.0030$ & $0.672/05 $ \\
 \hline
 $80$& $\Xi bar^+$ & $1.538\pm0.0444 $ & $-0.130\pm0.0064$& $0.490/04$ \\
 \hline
 $158$&$\Xi^-$ & $1.491\pm0.0268 $  & $-0.030\pm0.0031$ & $1.748/06 $ \\
 \hline
 $158$& $\Xi bar^+$ & $1.604\pm0.0336 $ & $-0.070\pm0.0031$& $0.621/05$ \\
 \hline
\end{tabular}
\end{small}
\end{center}
%\end{table}

%\begin{table}
\begin{center}
\begin{small}
\caption{Values of different parameters for production of $\Lambda$
and $\Lambda$bar in C+C and Si+Si collisions at 158A GeV(for $\beta$
= 0) for both +ve and -ve rapidities.[Reference Fig. No.13 \& 14]}
\begin{tabular}{|c|c|c|c|c|}\hline
 $Collision$&$Production$ & $C_3$ & $\gamma$ & $\frac{\chi^2}{ndf}$\\
 \hline
$C+C$& $\Lambda$ & $0.240\pm0.0027 $  & $0.033 \pm0.0051$ & $1.265/05$ \\
 \hline
 $C+C$& $\Lambda bar$ & $0.066 \pm0.0001 $ & $-0.075 \pm0.0015 $& $0.586/07 $ \\
 \hline
 $Si+Si$&$\Lambda$ & $0.841\pm0.0029 $  & $0.019 \pm0.0013$ & $4.757/07$ \\
 \hline
 $Si+Si$& $\Lambda bar$ & $0.163 \pm0.0014 $ & $-0.054 \pm0.0031 $& $0.997/09 $ \\
 \hline
 \end{tabular}
\end{small}
\end{center}
\end{table}

\newpage
\begin{figure}
\subfigure[]{
\begin{minipage}{.5\textwidth}
\centering
\includegraphics[width=2.5in]{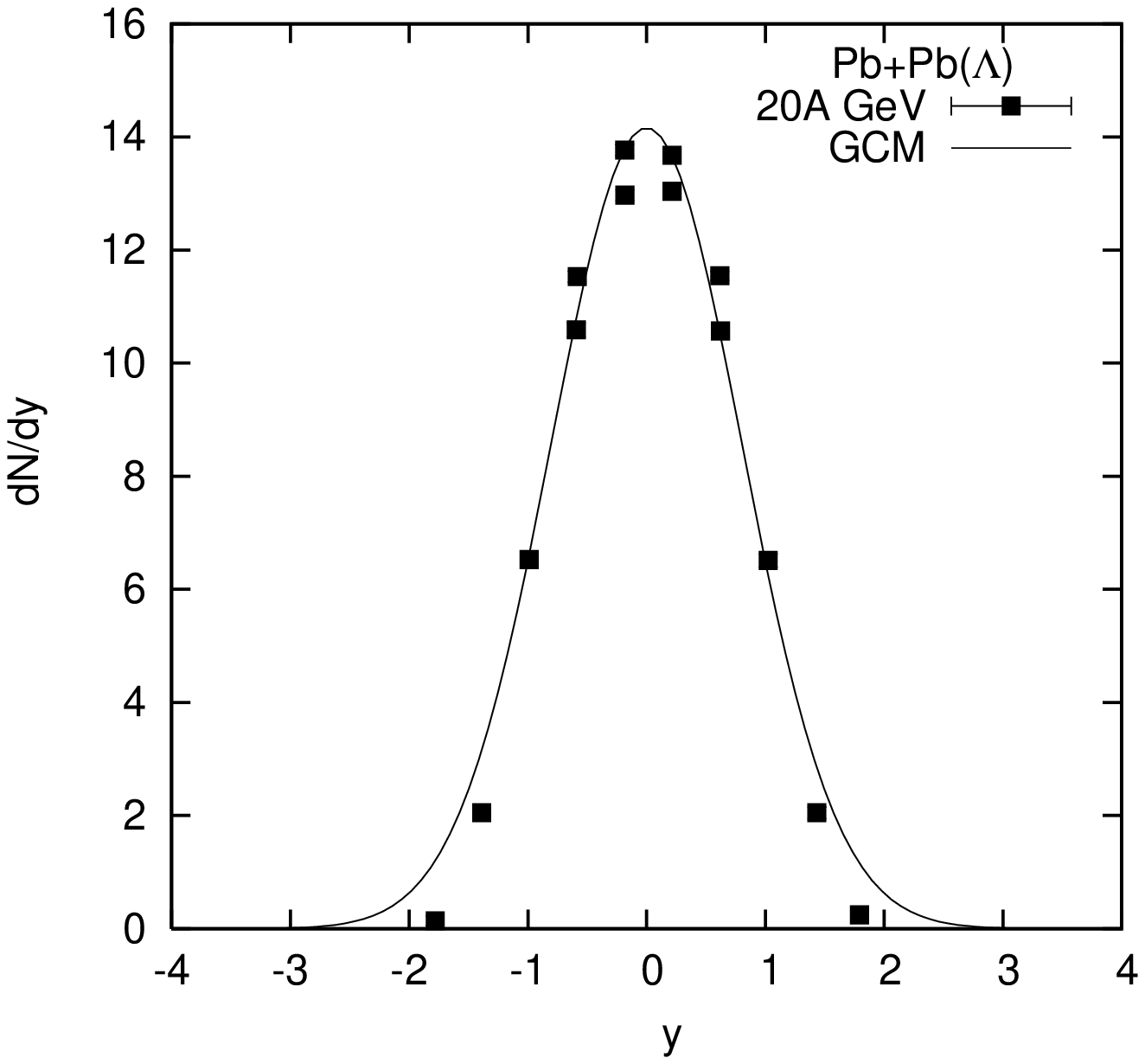}
\end{minipage}}%
\subfigure[]{
\begin{minipage}{.5\textwidth}
\centering
 \includegraphics[width=2.5in]{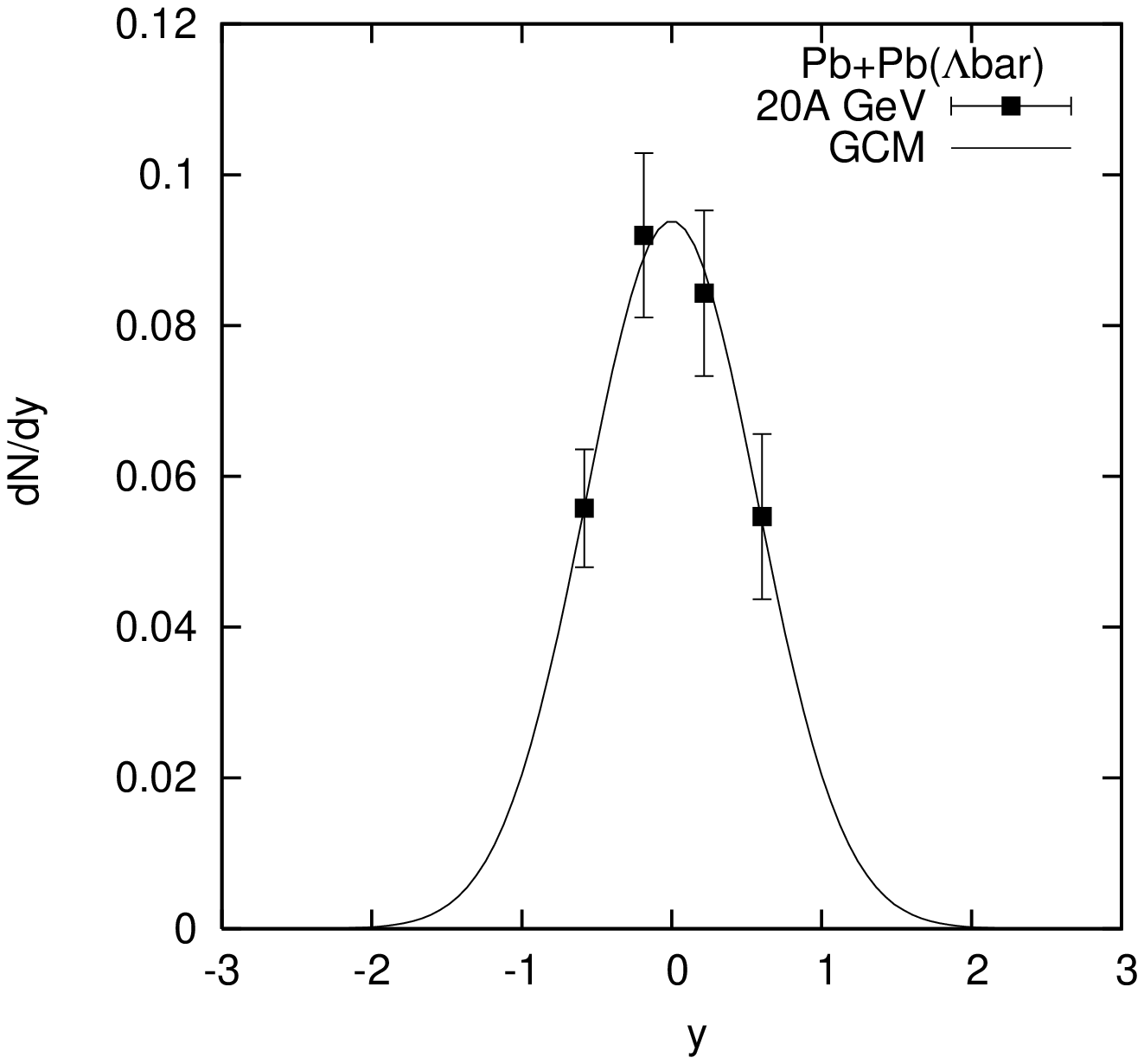}
 \end{minipage}}%
\caption{Plot of $\frac{dN}{dy}$ vs. y for $\Lambda$ and $\Lambda$bar produced in Pb+Pb collisions
at 20A GeV for $\beta$=0. The different experimental points are taken from {\cite{Alt1}} and the parameter
values are taken from Table 2. The solid curve provide the GCM-based results.}
%\end{figure}

%\begin{figure}
\subfigure[]{
\begin{minipage}{.5\textwidth}
\centering
\includegraphics[width=2.5in]{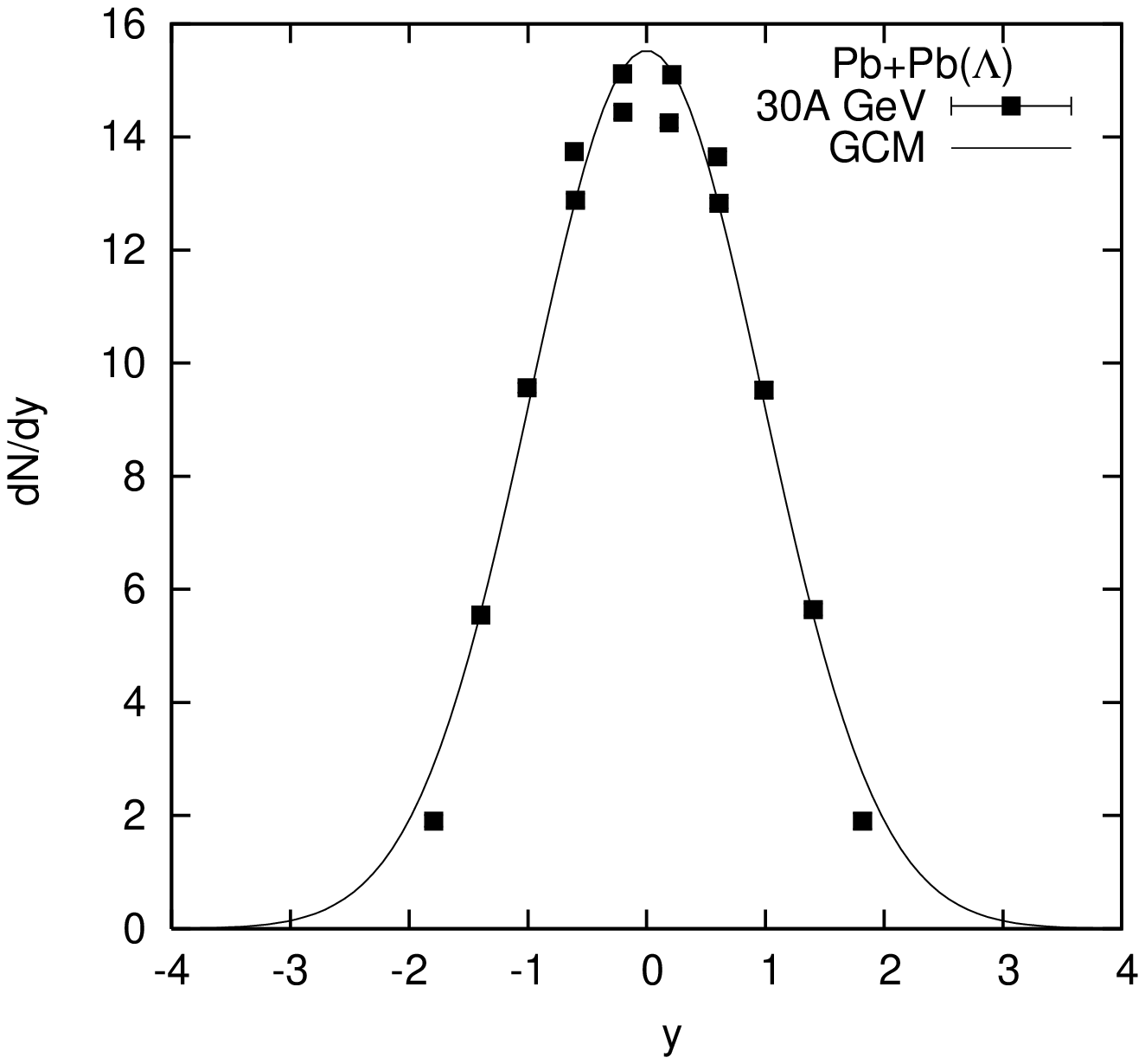}
\end{minipage}}%
\subfigure[]{
\begin{minipage}{.5\textwidth}
\centering
 \includegraphics[width=2.5in]{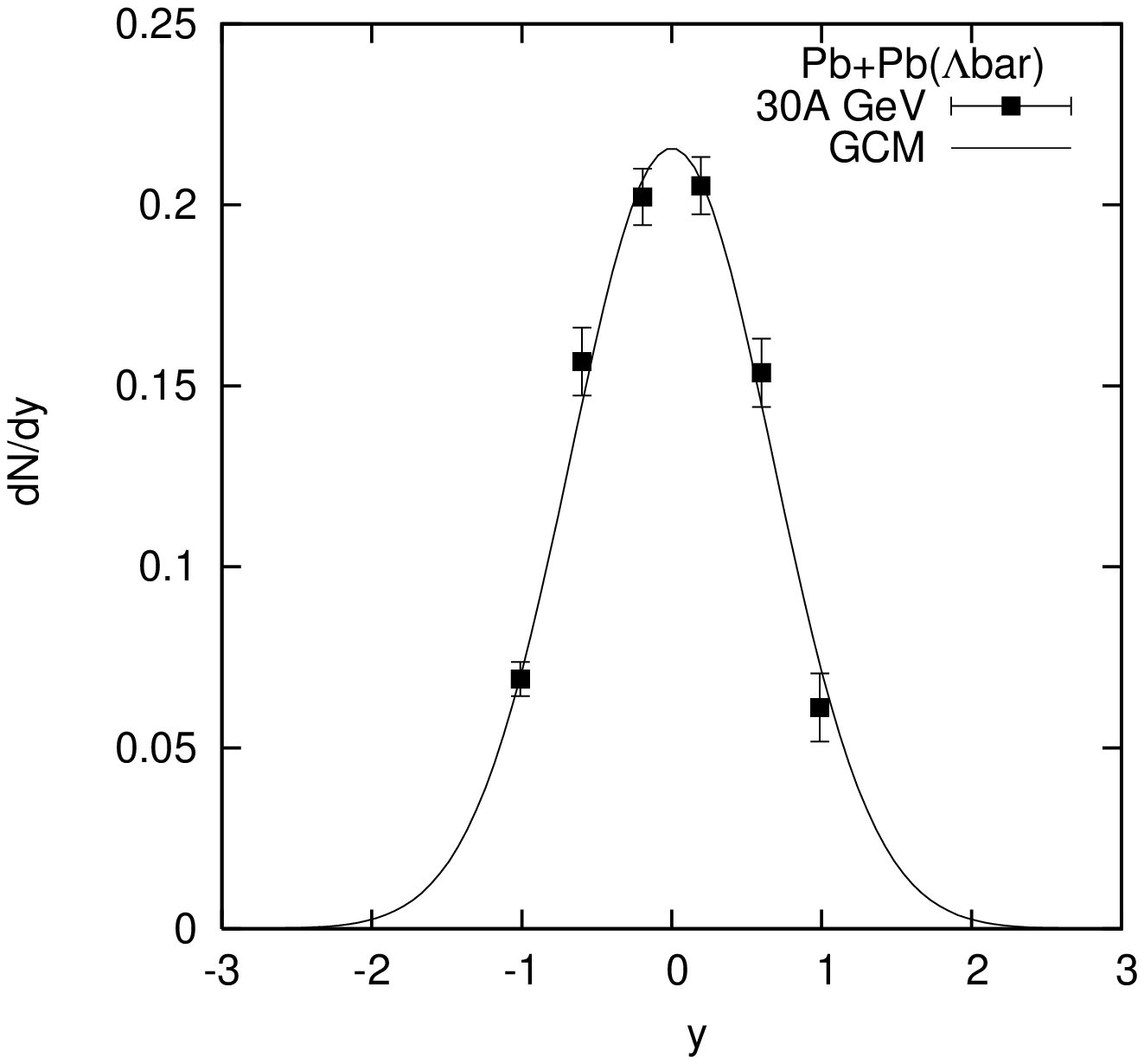}
 \end{minipage}}%
\caption{Rapidity spectra for $\Lambda$ and $\Lambda$bar produced in
Pb+Pb collisions at 30A GeV for $\beta$=0. The different
experimental points are taken from {\cite{Alt1}} and the parameter
values are taken from Table 2. The solid curve provide the GCM-based
results.}
\end{figure}

\begin{figure}
\subfigure[]{
\begin{minipage}{.5\textwidth}
\centering
\includegraphics[width=2.5in]{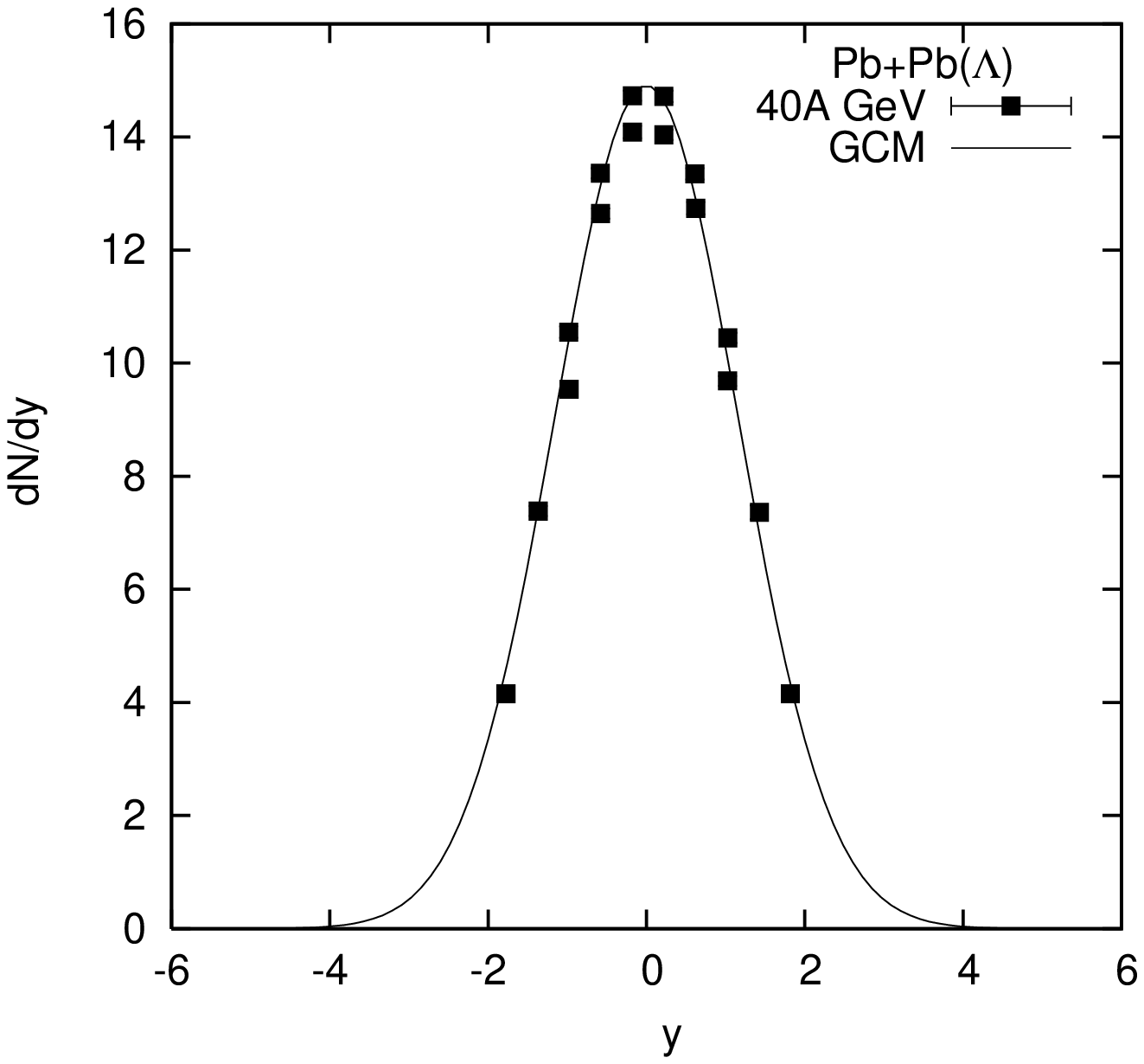}
\end{minipage}}%
\subfigure[]{
\begin{minipage}{.5\textwidth}
\centering
 \includegraphics[width=2.5in]{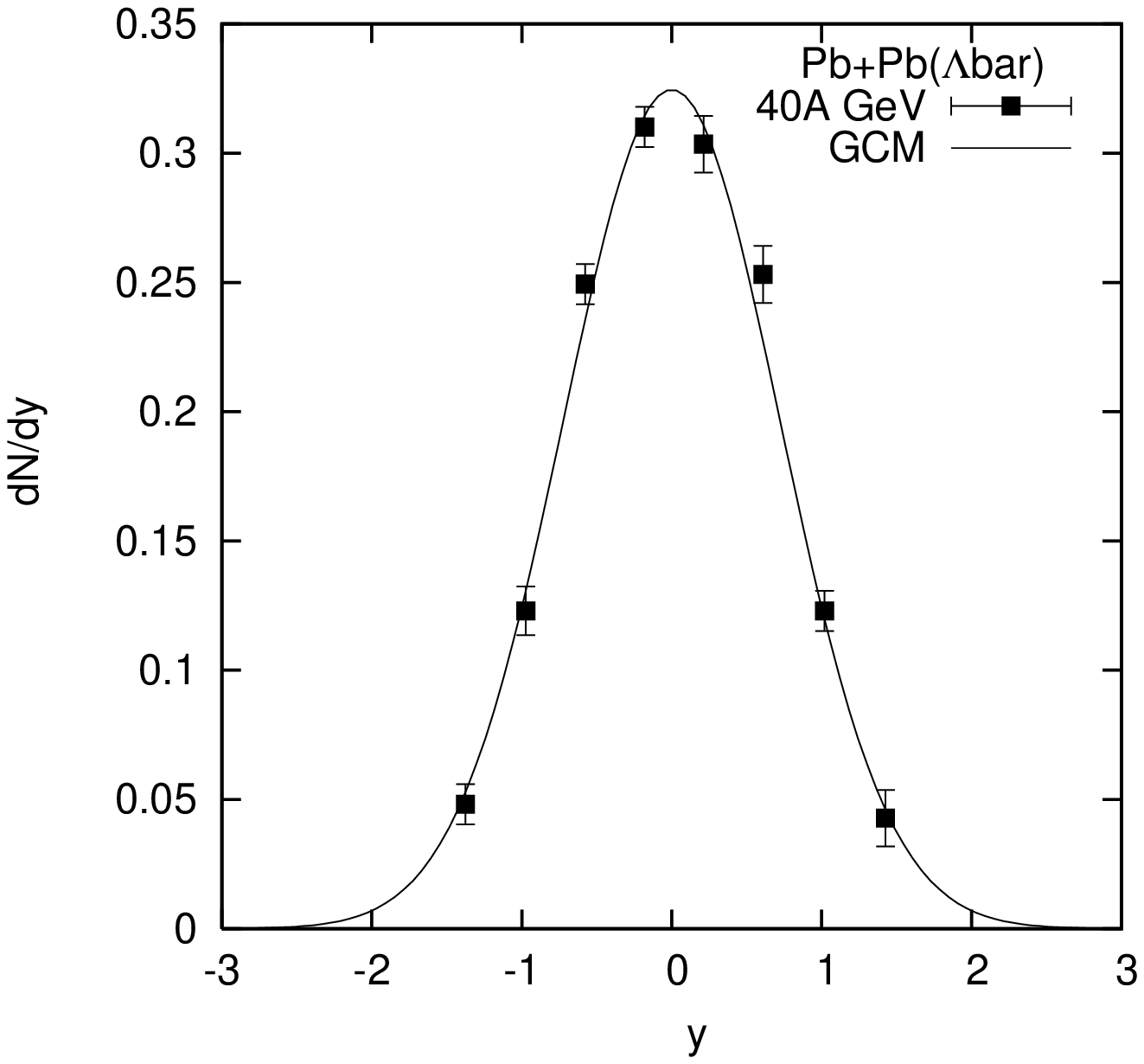}
 \end{minipage}}%
\caption{Plot of $\frac{dN}{dy}$ vs. y for $\Lambda$ and $\Lambda$bar produced in Pb+Pb collisions
at 40A GeV for $\beta$=0. The different experimental points are taken from {\cite{Alt1}} and the parameter
values are taken from Table 2. The solid curve provide the GCM-based results.}
%\end{figure}

%\begin{figure}
\subfigure[]{
\begin{minipage}{.5\textwidth}
\centering
\includegraphics[width=2.5in]{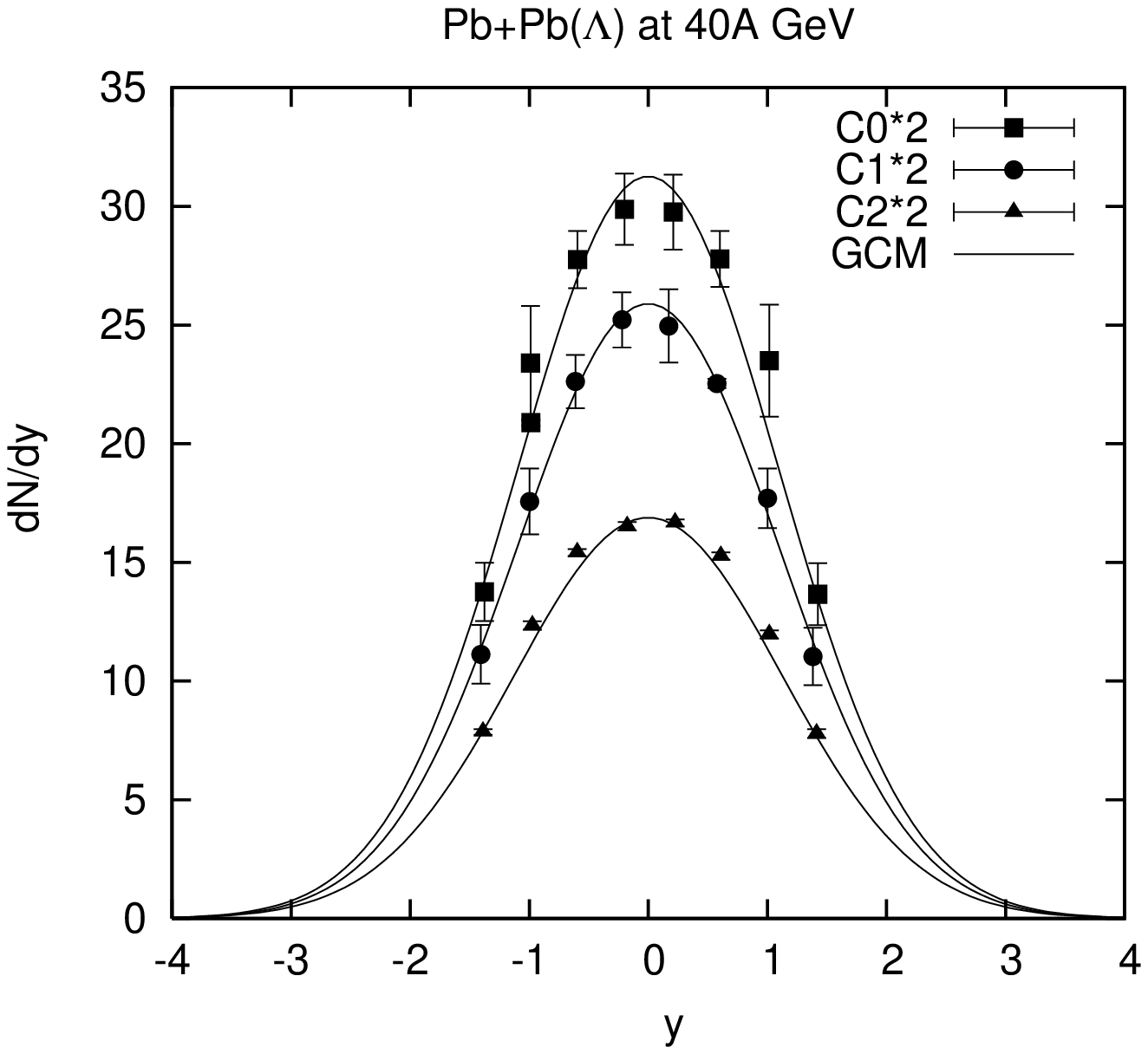}
\end{minipage}}%
\subfigure[]{
\begin{minipage}{.5\textwidth}
\centering
 \includegraphics[width=2.5in]{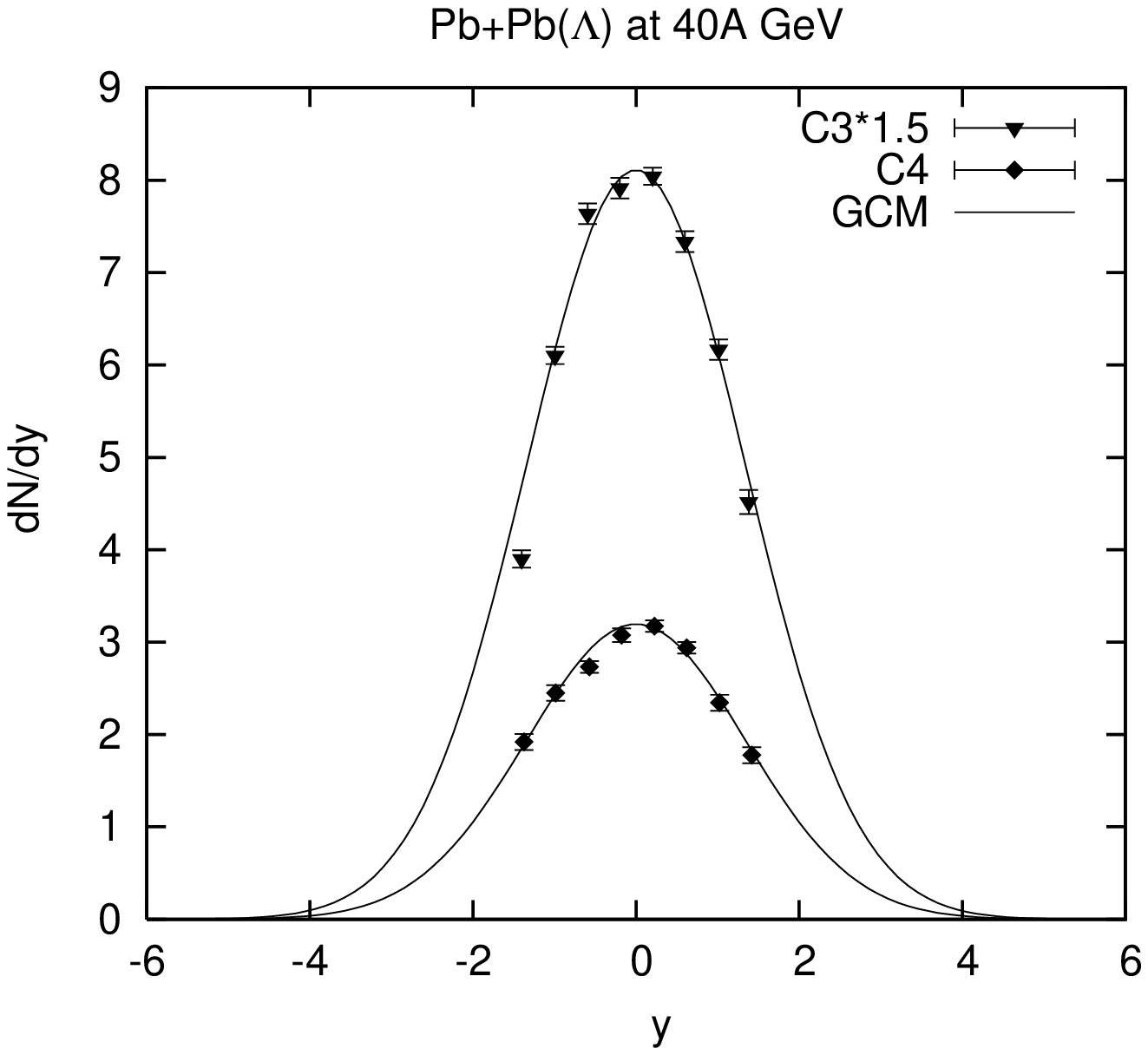}
 \end{minipage}}%
\caption{The rapidity spectra of $\Lambda$ for Pb+Pb collisions at
40A GeV for the five different centrality bins C0-C4 for $\beta$=0.
The experimental points are taken from \cite{Anticic1} and the
parameter values are taken from Table 3. The solid curves provide
the GCM-based results. }
\end{figure}

\begin{figure}
\subfigure[]{
\begin{minipage}{.5\textwidth}
\centering
\includegraphics[width=2.5in]{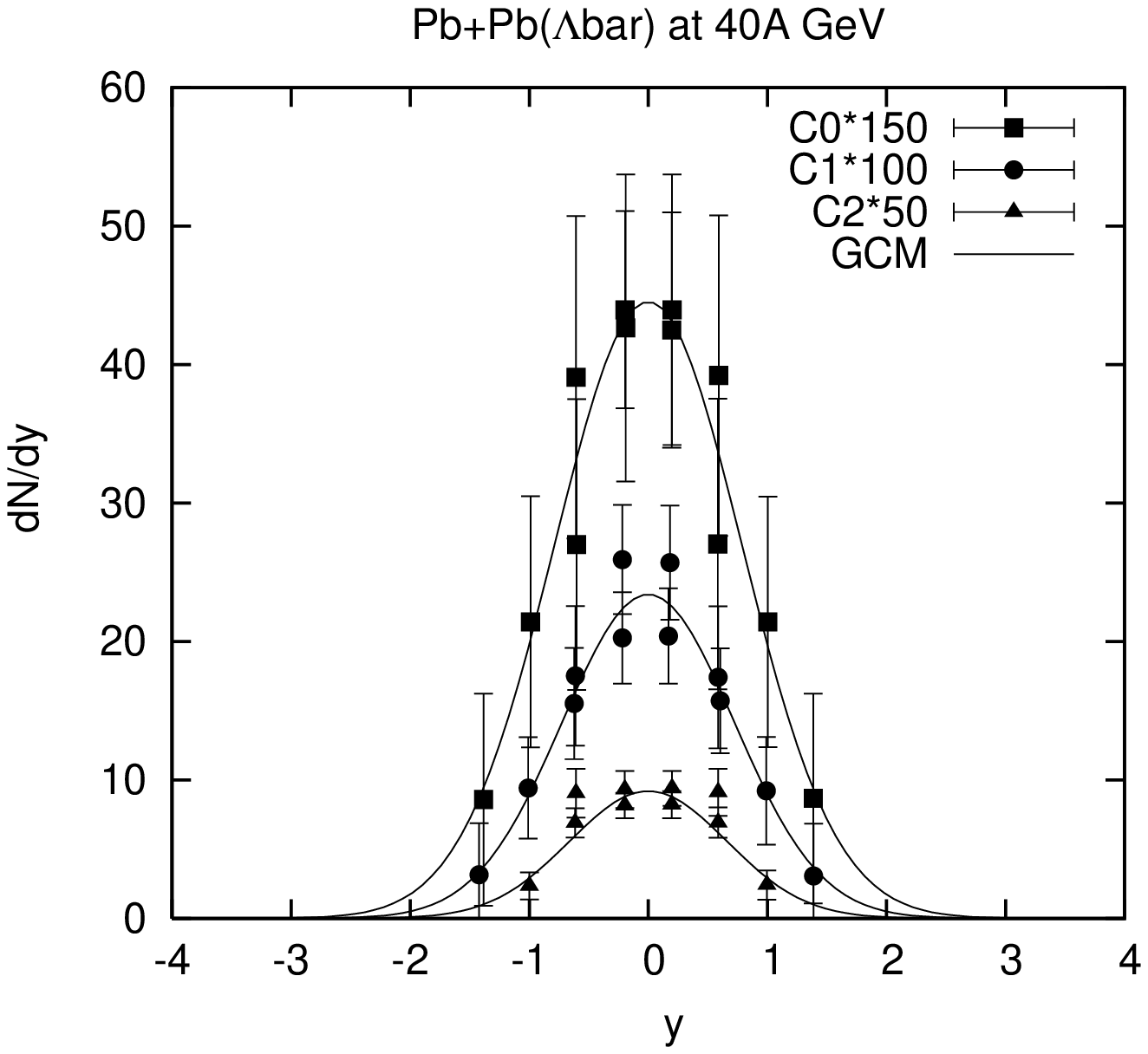}
\end{minipage}}%
\subfigure[]{
\begin{minipage}{.5\textwidth}
\centering
 \includegraphics[width=2.5in]{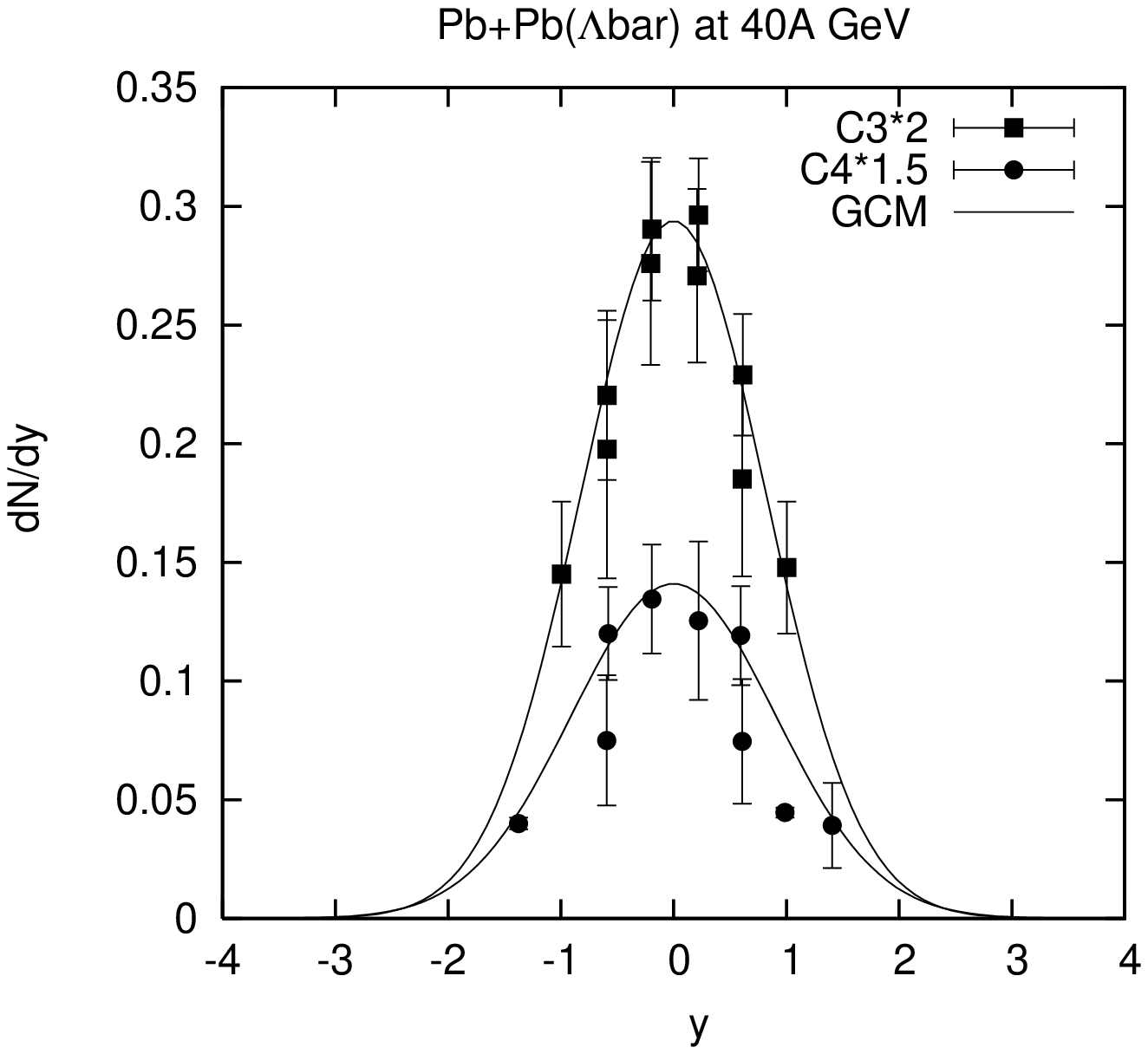}
 \end{minipage}}%
\caption{The rapidity spectra of $\Lambda$bar for Pb+Pb collisions at 40A GeV for the five
different centrality bins C0-C4 for $\beta$=0. The experimental points are taken from \cite{Anticic1} and the parameter
values are taken from Table 4. The solid curves provide the GCM-based results.}
\end{figure}

\begin{figure}
\subfigure[]{
\begin{minipage}{.5\textwidth}
\centering
\includegraphics[width=2.5in]{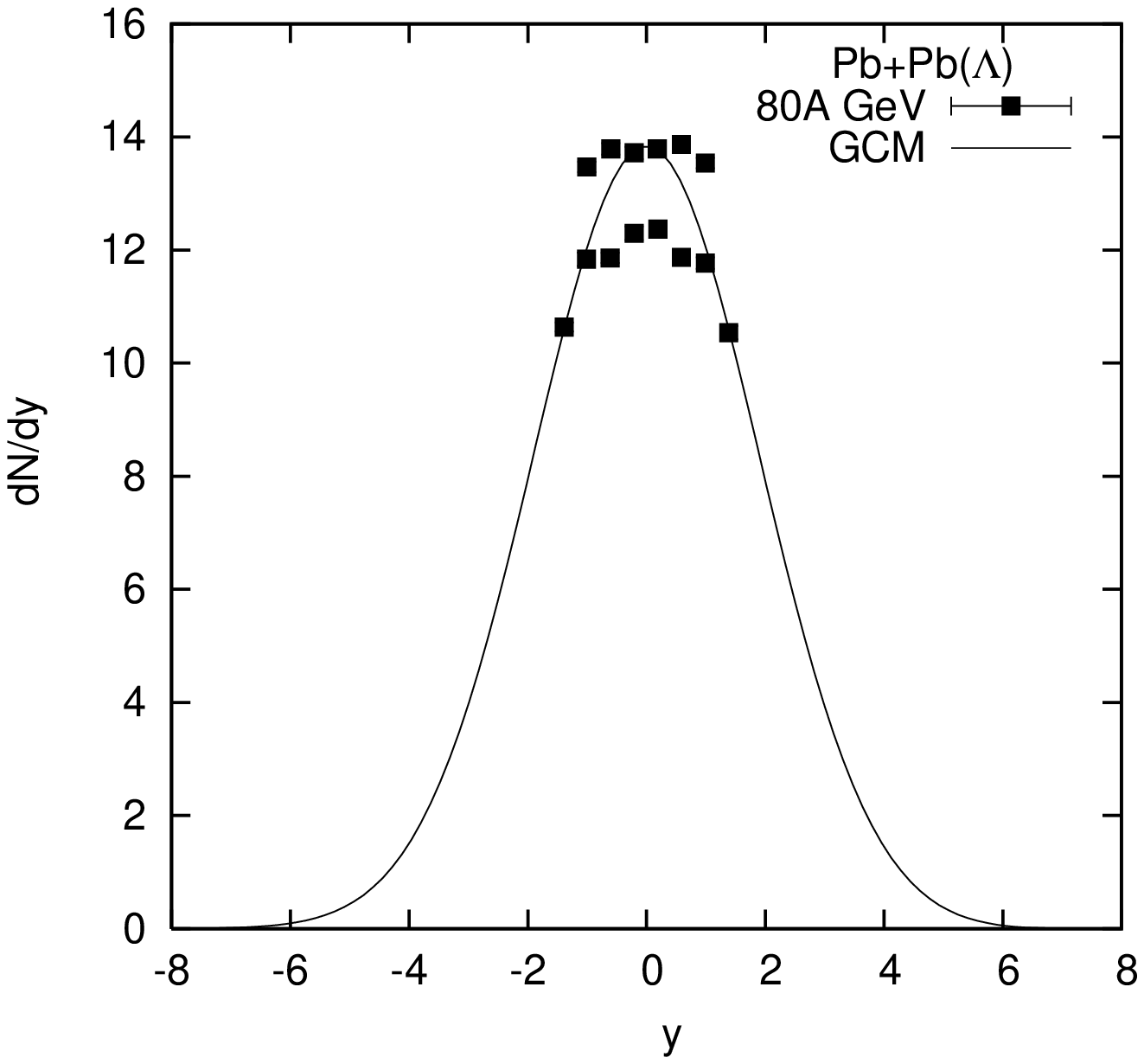}
\setcaptionwidth{2.6in}
\end{minipage}}%
\subfigure[]{
\begin{minipage}{0.5\textwidth}
\centering
 \includegraphics[width=2.5in]{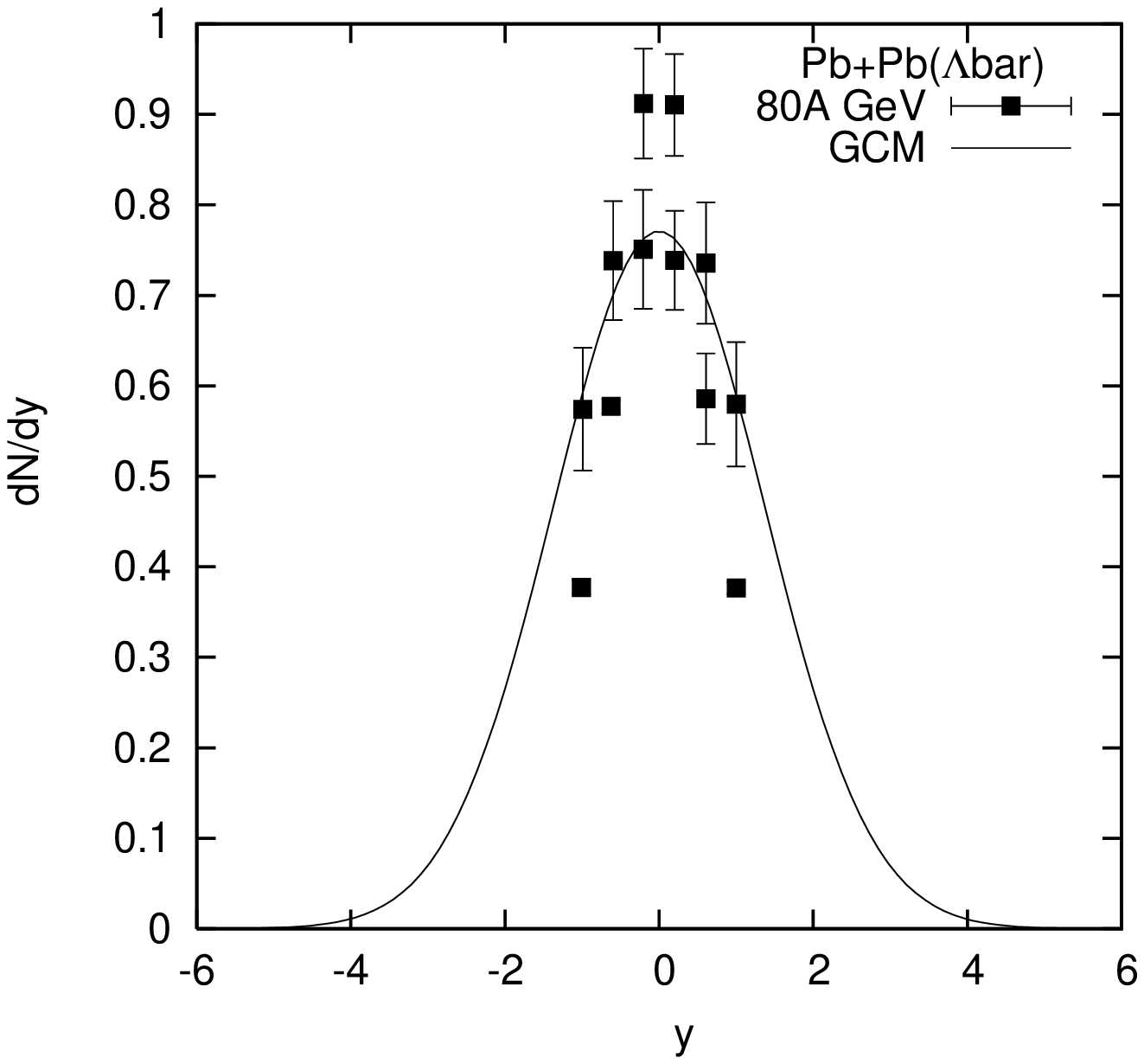}
 \end{minipage}}%
\vspace{0.01in} \subfigure[]{
\begin{minipage}{0.5\textwidth}
\centering
\includegraphics[width=2.5in]{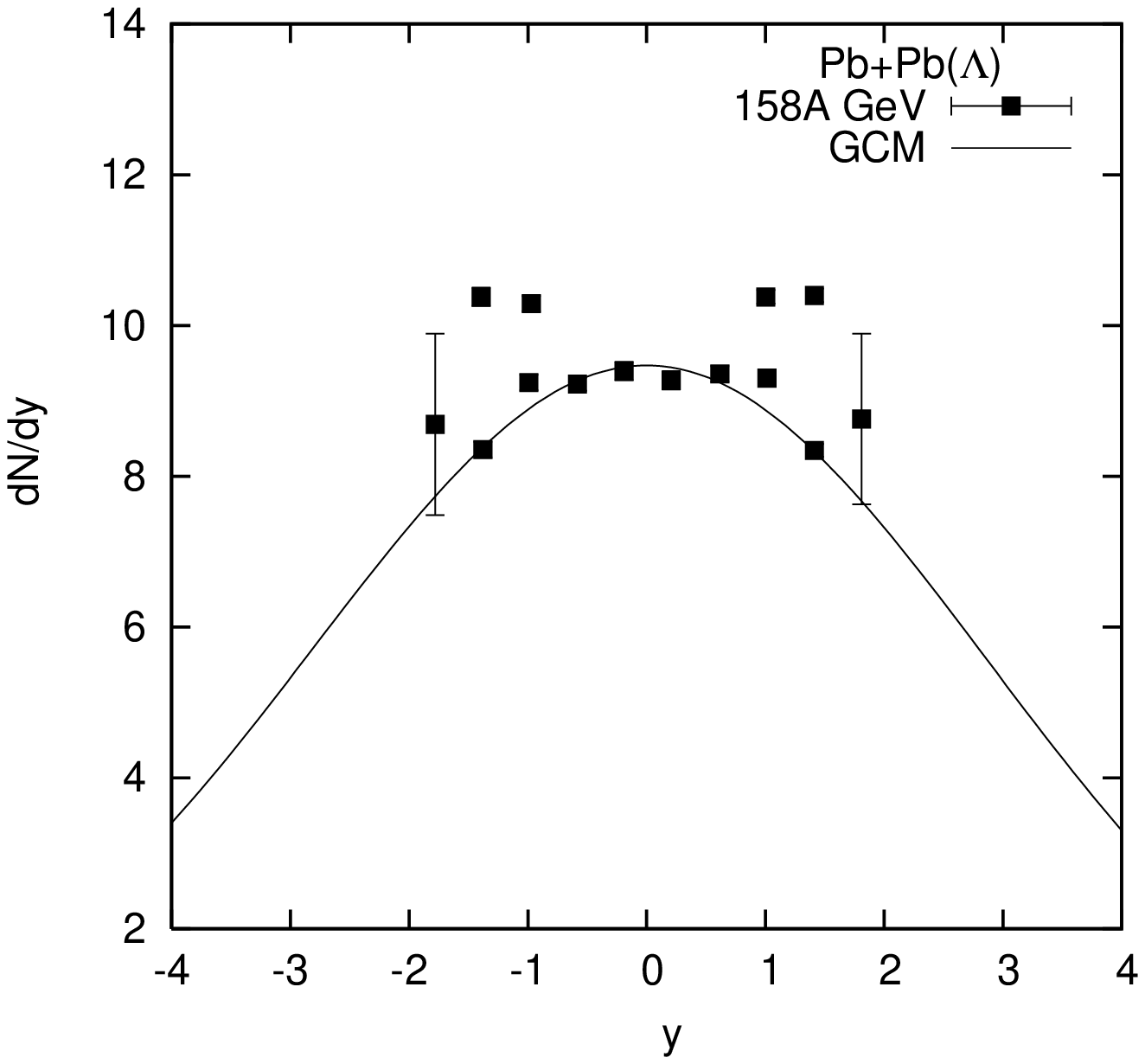}
\end{minipage}}%
\subfigure[]{
\begin{minipage}{.5\textwidth}
\centering
 \includegraphics[width=2.5in]{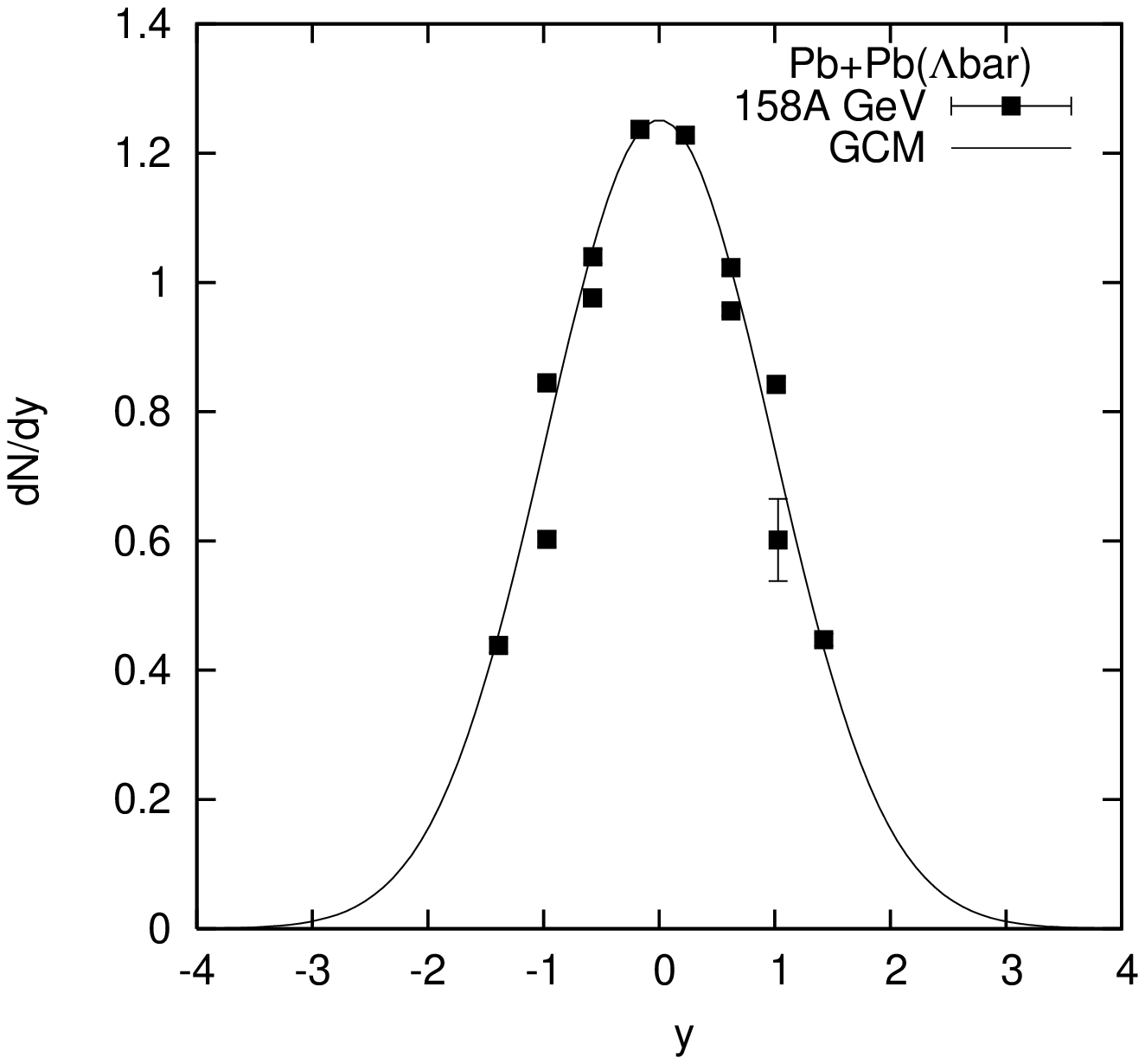}
 \end{minipage}}%
\caption{Rapidity spectra for $\Lambda$ and $\Lambda$bar produced in Pb+Pb collisions
at 80A GeV [Fig.6(a) and Fig.6(b)] and 158A GeV [Fig.6(c) and Fig.6(d)] for $\beta$=0.
The different experimental points are taken from {\cite{Alt1}} and the parameter
values are taken from Table 5. The solid curve provide the GCM-based results.}
\end{figure}

\begin{figure}
\subfigure[]{
\begin{minipage}{.5\textwidth}
\centering
\includegraphics[width=2.5in]{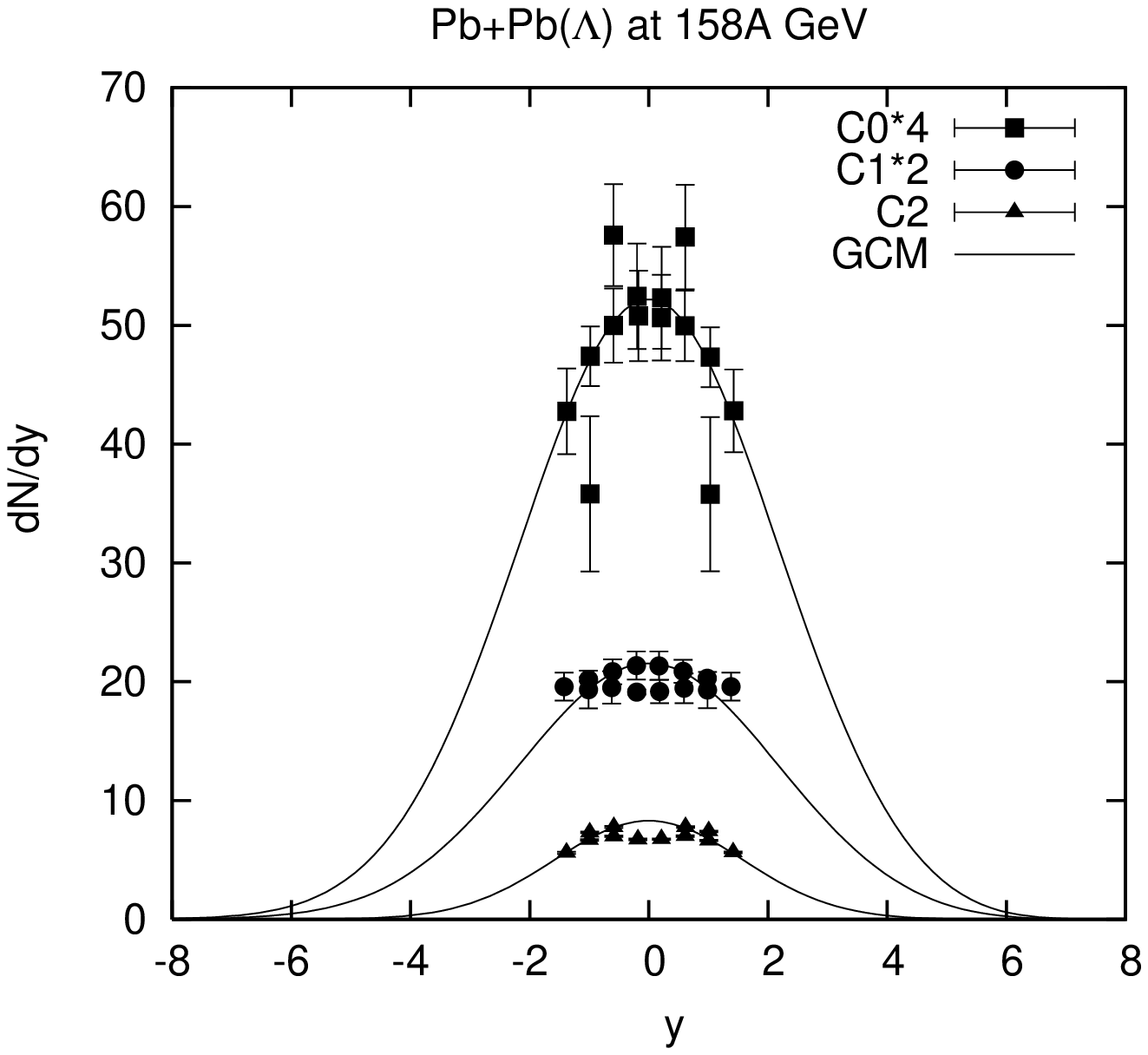}
\end{minipage}}%
\subfigure[]{
\begin{minipage}{.5\textwidth}
\centering
 \includegraphics[width=2.5in]{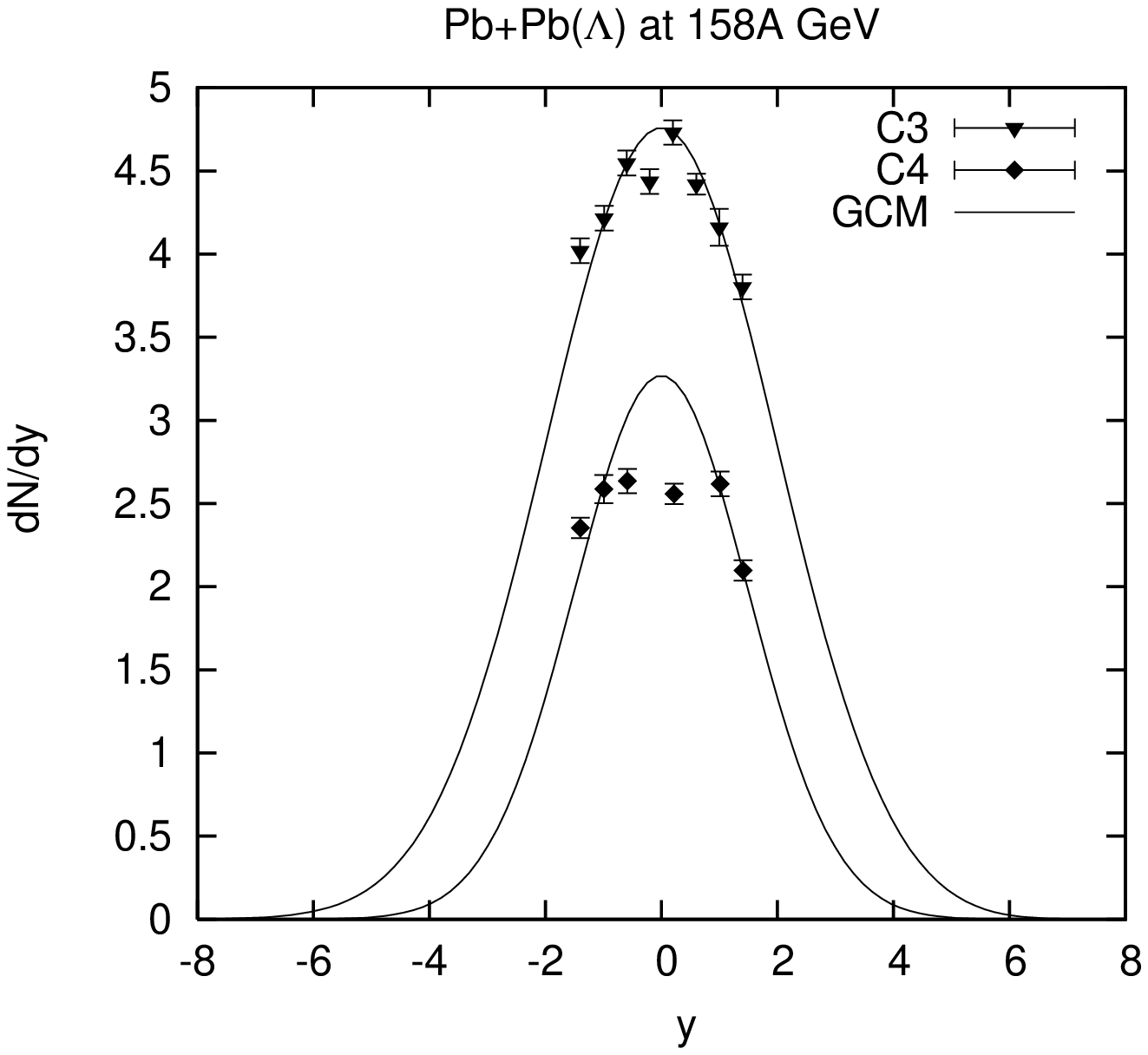}
 \end{minipage}}%
\caption{The rapidity spectra of $\Lambda$ for Pb+Pb collisions at 158A GeV for the five
different centrality bins C0-C4 for $\beta$=0. The experimental points are taken from \cite{Anticic1} and the parameter
values are taken from Table 6. The solid curves provide the GCM-based results.}
\end{figure}

\begin{figure}
\subfigure[]{
\begin{minipage}{.5\textwidth}
\centering
\includegraphics[width=2.5in]{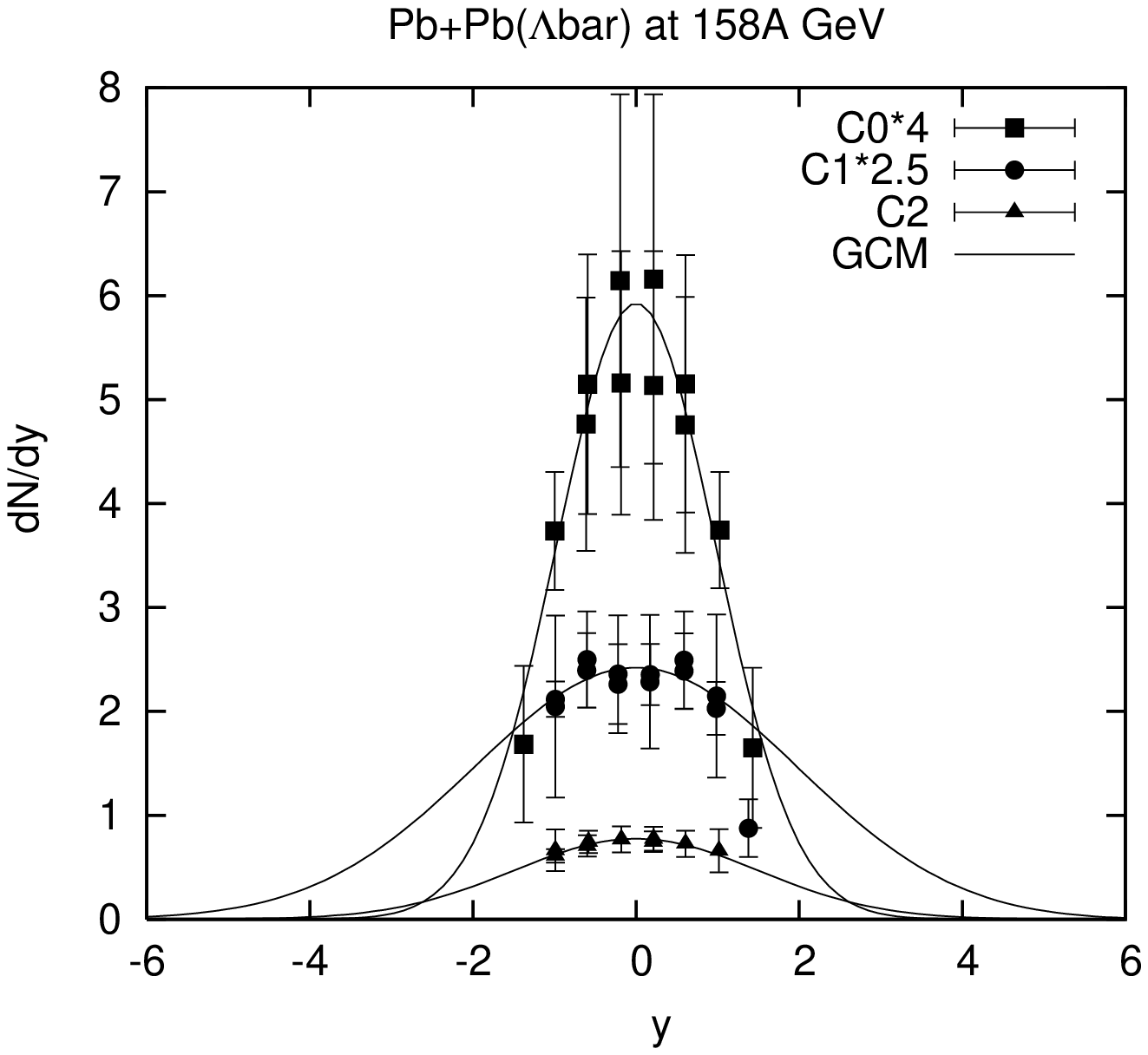}
\end{minipage}}%
\subfigure[]{
\begin{minipage}{.5\textwidth}
\centering
 \includegraphics[width=2.5in]{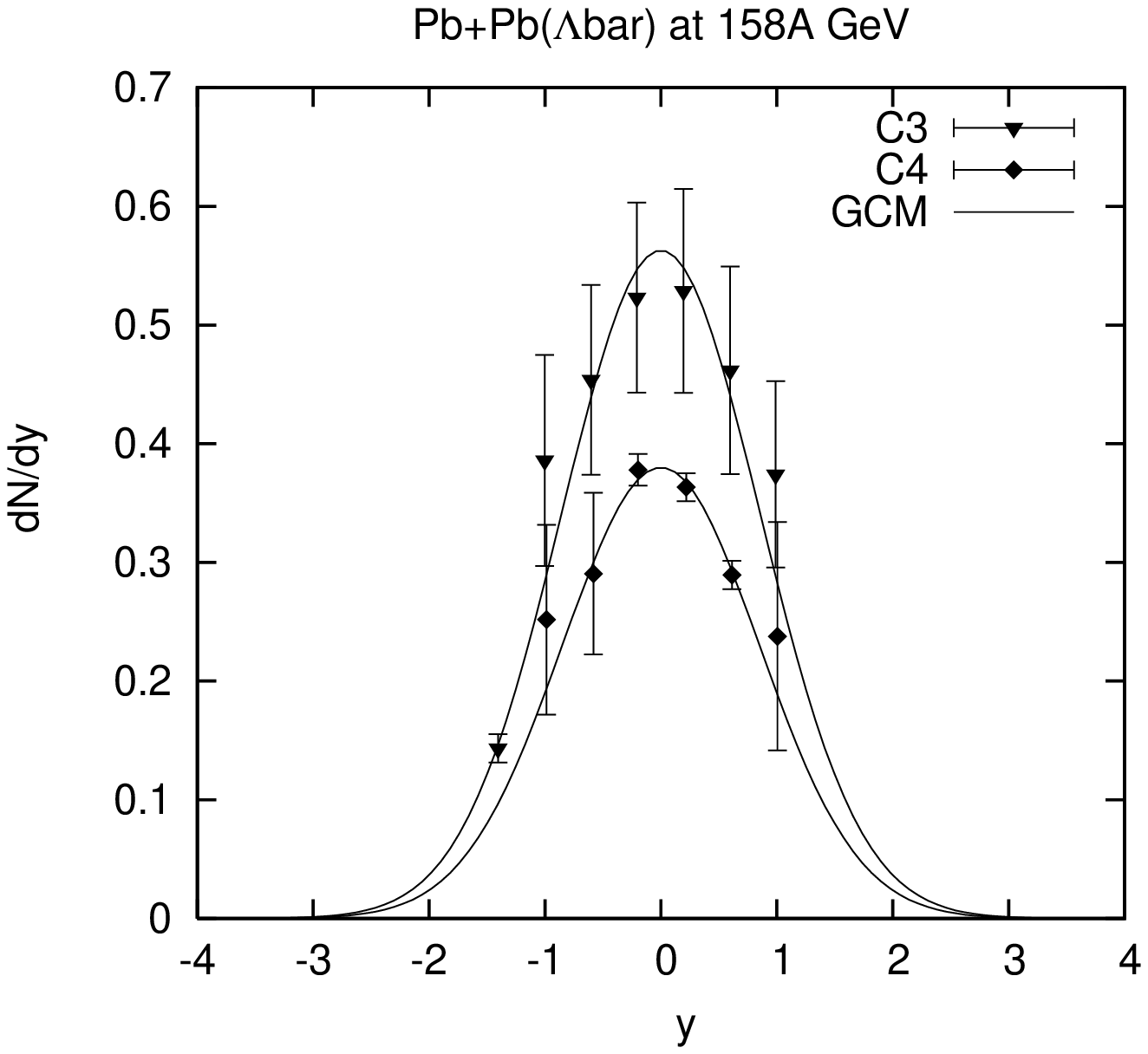}
 \end{minipage}}%
\caption{The rapidity spectra of $\Lambda$bar for Pb+Pb collisions at 158A GeV for the five
different centrality bins C0-C4 for $\beta$=0. The experimental points are taken from \cite{Anticic1} and the parameter
values are taken from Table 7. The solid curves provide the GCM-based results.}
%\end{figure}

%\begin{figure}
\centering
\includegraphics[width=2.5in]{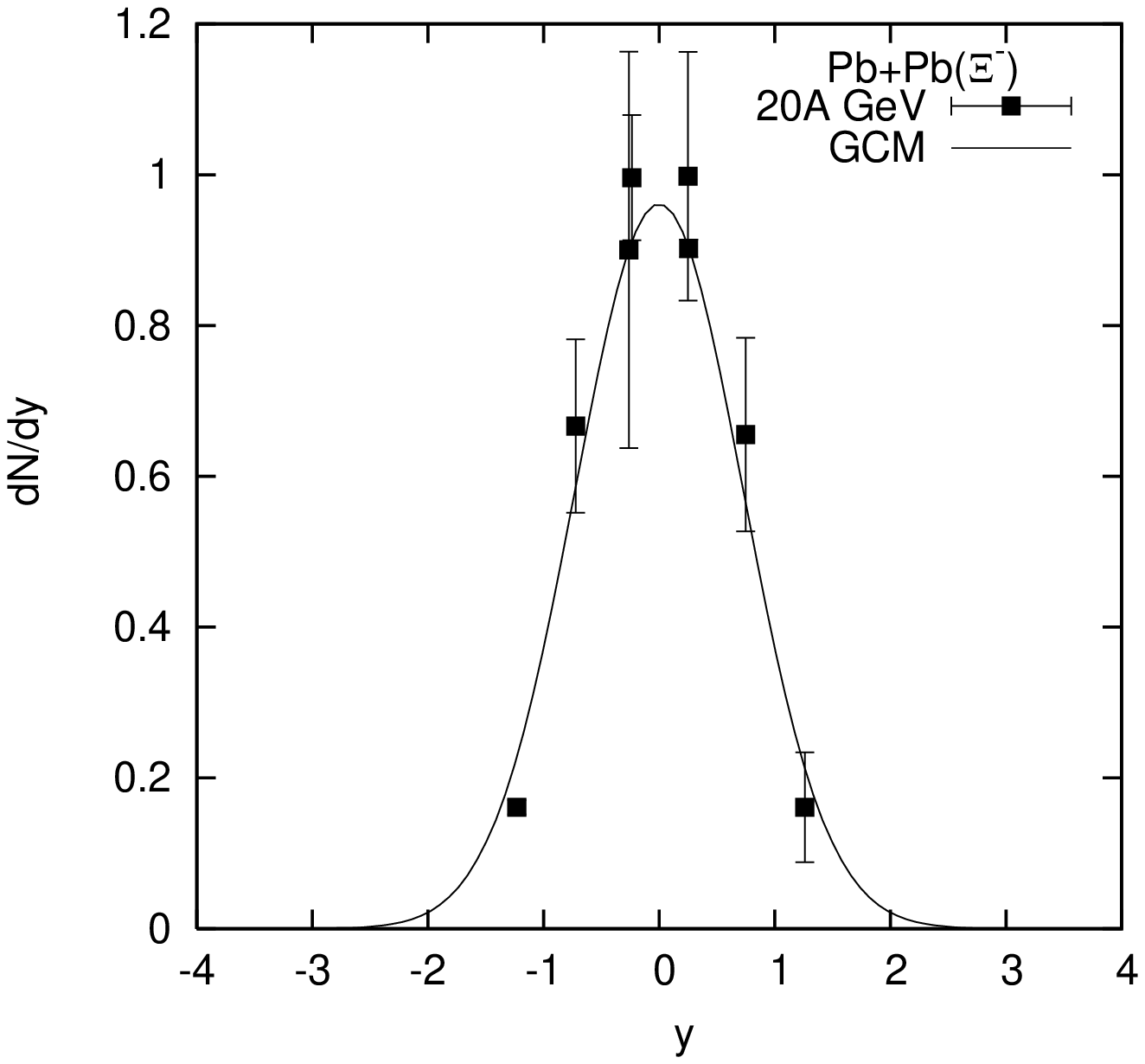}
\caption{Rapidity spectra for $\Xi^-$ produced in Pb+Pb collisions
at 20A GeV for $\beta$=0. The different experimental points are
taken from {\cite{Alt1}} and the parameter values are taken from
Table 8. The solid curve provide the GCM-based results.}
\end{figure}

\begin{figure}
\subfigure[]{
\begin{minipage}{.5\textwidth}
\centering
\includegraphics[width=2.5in]{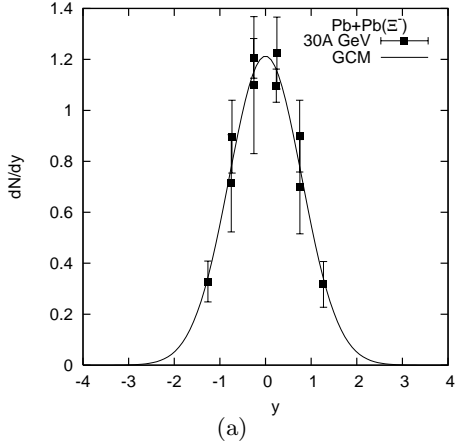}
\setcaptionwidth{2.6in}
\end{minipage}}%
\subfigure[]{
\begin{minipage}{0.5\textwidth}
\centering
 \includegraphics[width=2.5in]{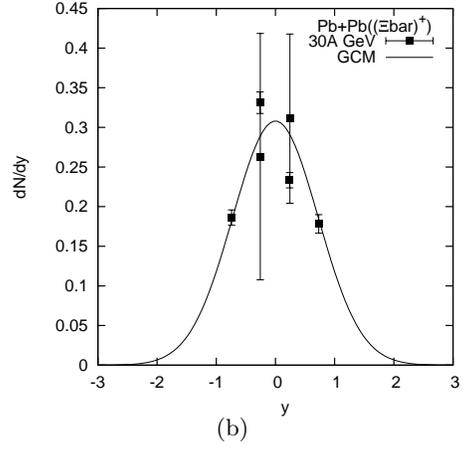}
 \end{minipage}}%
\vspace{0.01in} \subfigure[]{
\begin{minipage}{0.5\textwidth}
\centering
\includegraphics[width=2.5in]{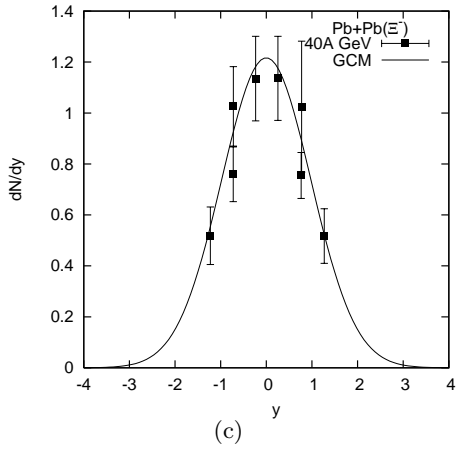}
\end{minipage}}%
\subfigure[]{
\begin{minipage}{.5\textwidth}
\centering
 \includegraphics[width=2.5in]{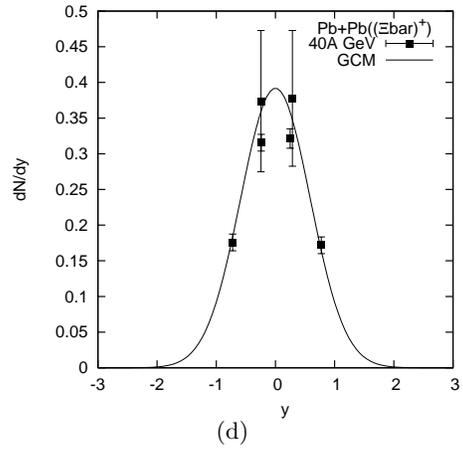}
 \end{minipage}}%
\caption{Plot of $\frac{dN}{dy}$ vs. y for $\Xi^-$ and $\Xi$bar$^+$ produced in Pb+Pb collisions
at 30A GeV [Fig.12(a) and Fig.12(b)] and 40A GeV [Fig.12(c) and Fig.12(d)] for $\beta$=0.
The different experimental points are taken from {\cite{Alt1}} and the parameter
values are taken from Table 8. The solid curve provide the GCM-based results.}
\end{figure}

\begin{figure}
\subfigure[]{
\begin{minipage}{.5\textwidth}
\centering
\includegraphics[width=2.5in]{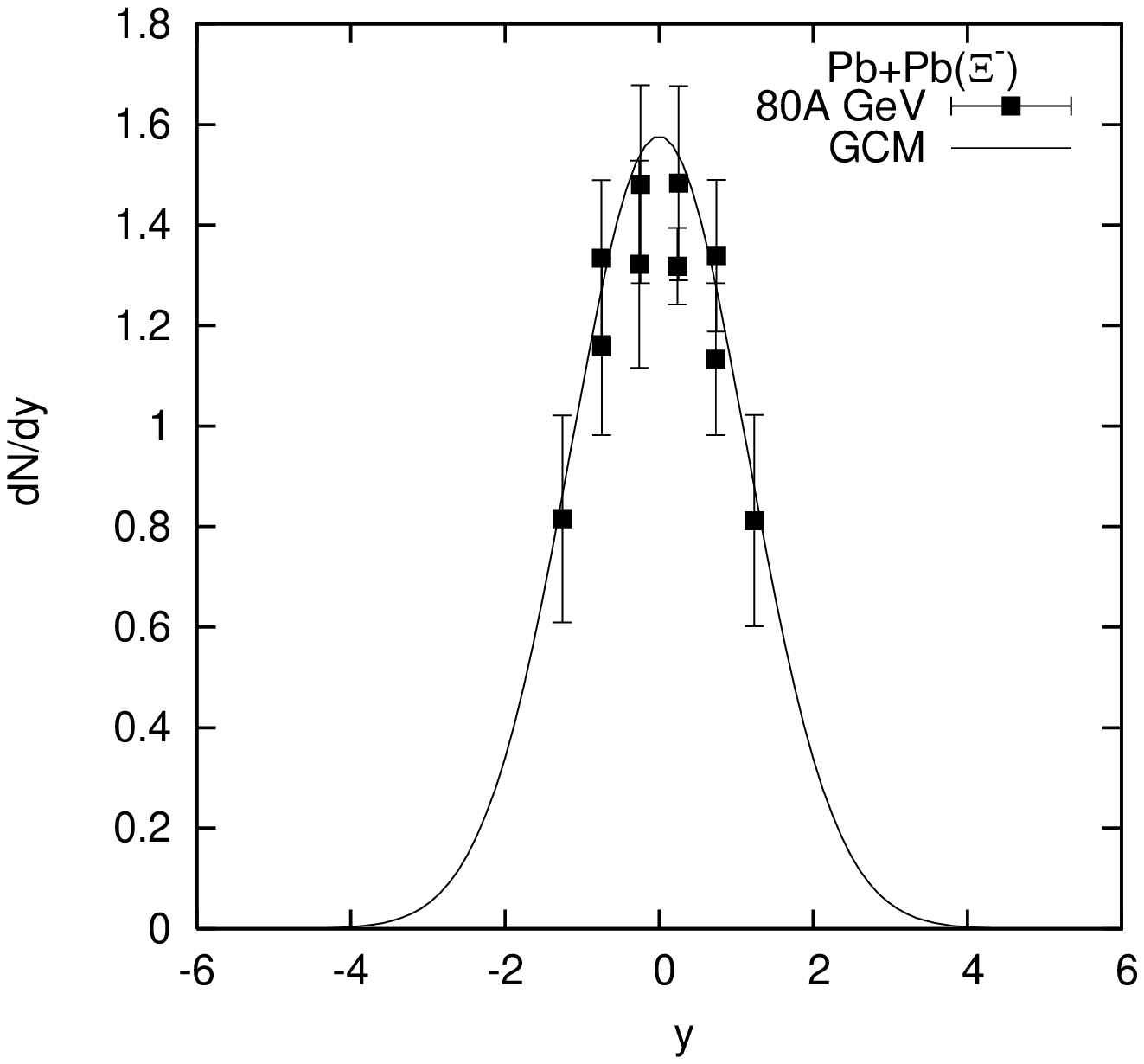}
\setcaptionwidth{2.6in}
\end{minipage}}%
\subfigure[]{
\begin{minipage}{0.5\textwidth}
\centering
 \includegraphics[width=2.5in]{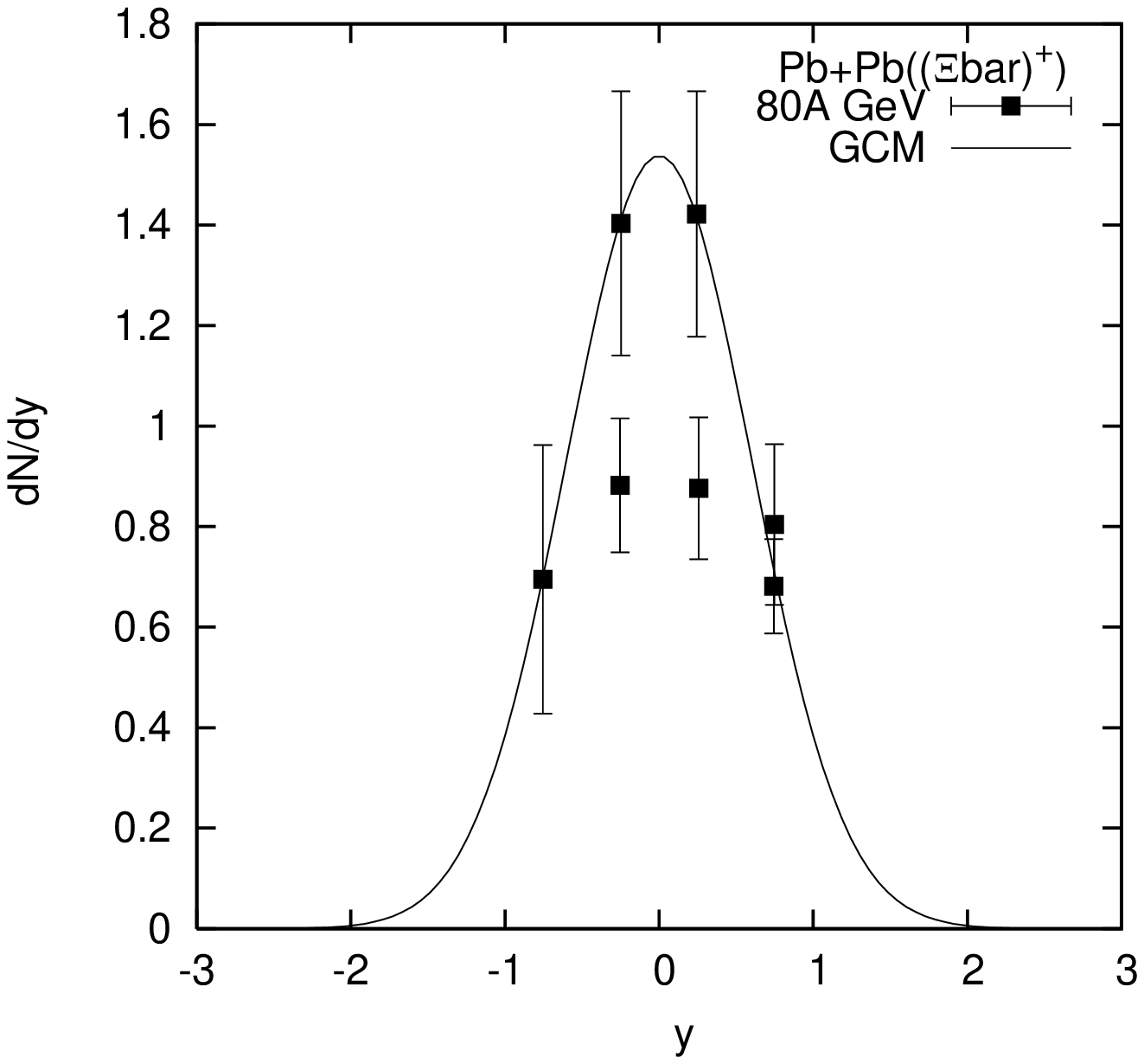}
 \end{minipage}}%
\vspace{0.01in} \subfigure[]{
\begin{minipage}{0.5\textwidth}
\centering
\includegraphics[width=2.5in]{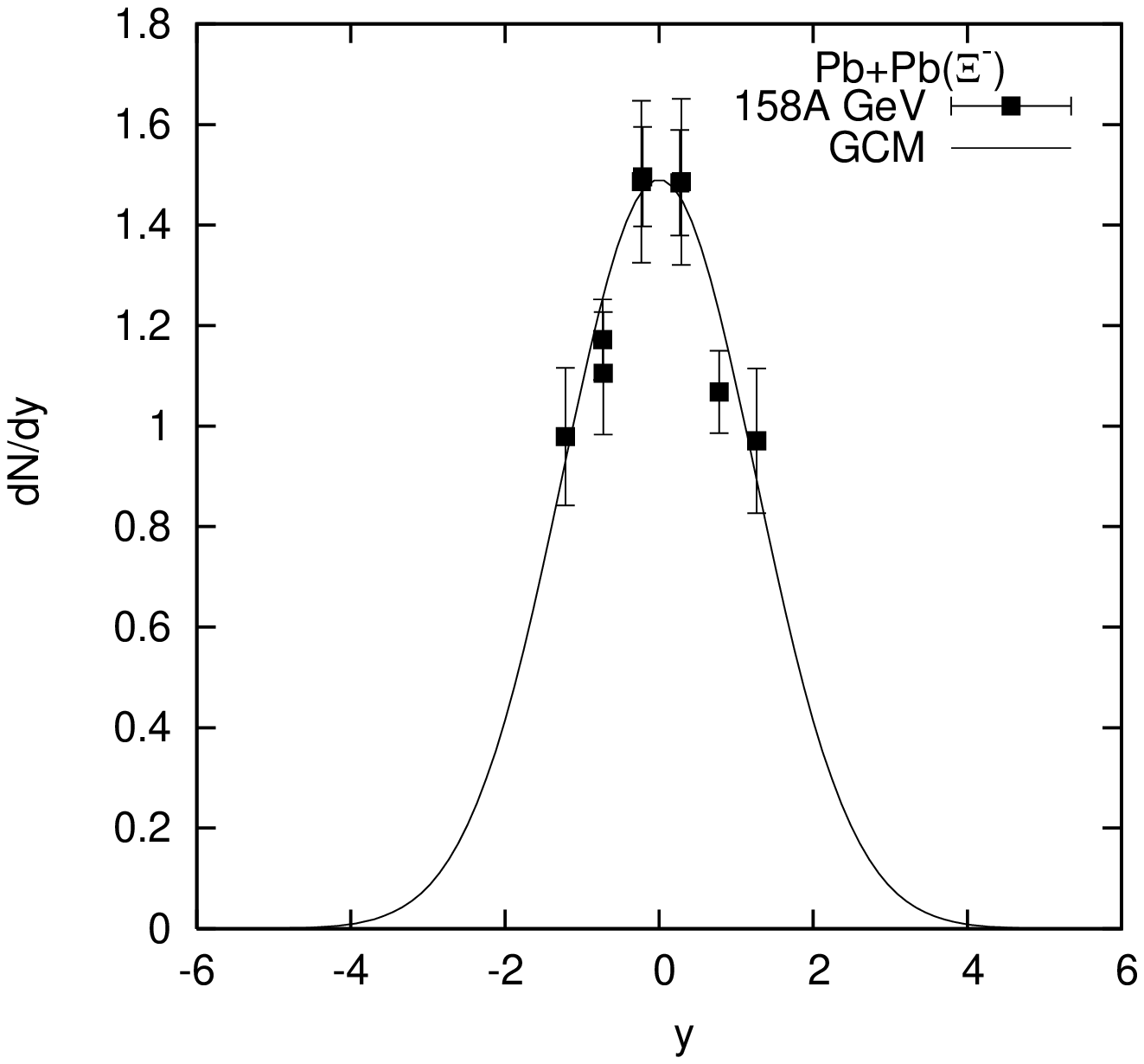}
\end{minipage}}%
\subfigure[]{
\begin{minipage}{.5\textwidth}
\centering
 \includegraphics[width=2.5in]{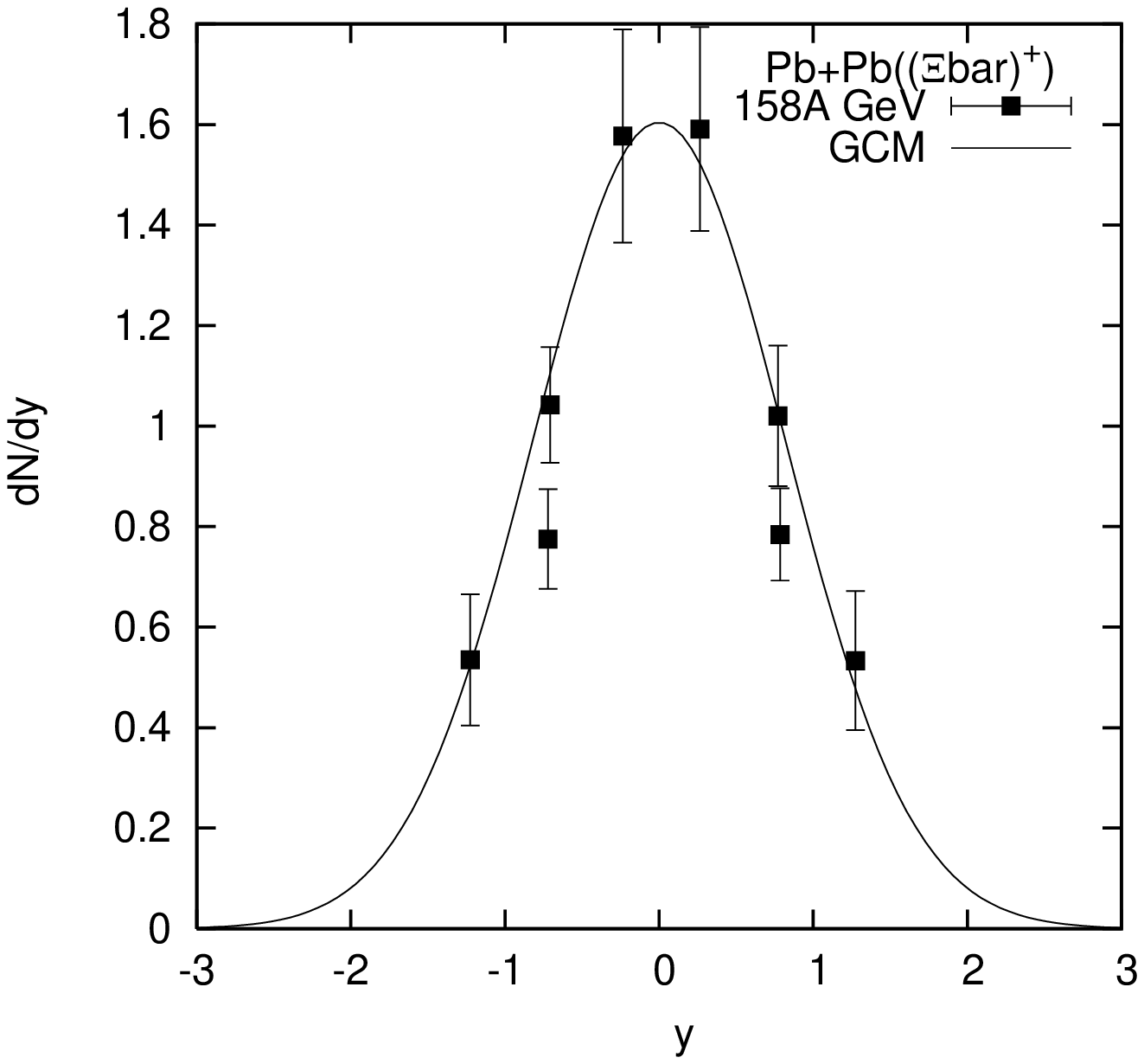}
 \end{minipage}}%
\caption{Rapidity spectra for $\Xi^-$ and $\Xi$bar$^+$ produced in Pb+Pb collisions
at 80A GeV [Fig.11(a) and Fig.11(b)] and 158A GeV [Fig.11(c) and Fig.11(d)] for $\beta$=0.
The different experimental points are taken from {\cite{Alt1}} and the parameter
values are taken from Table 9. The solid curve provide the GCM-based results.}
\end{figure}

\begin{figure}
\subfigure[]{
\begin{minipage}{.5\textwidth}
\centering
\includegraphics[width=2.5in]{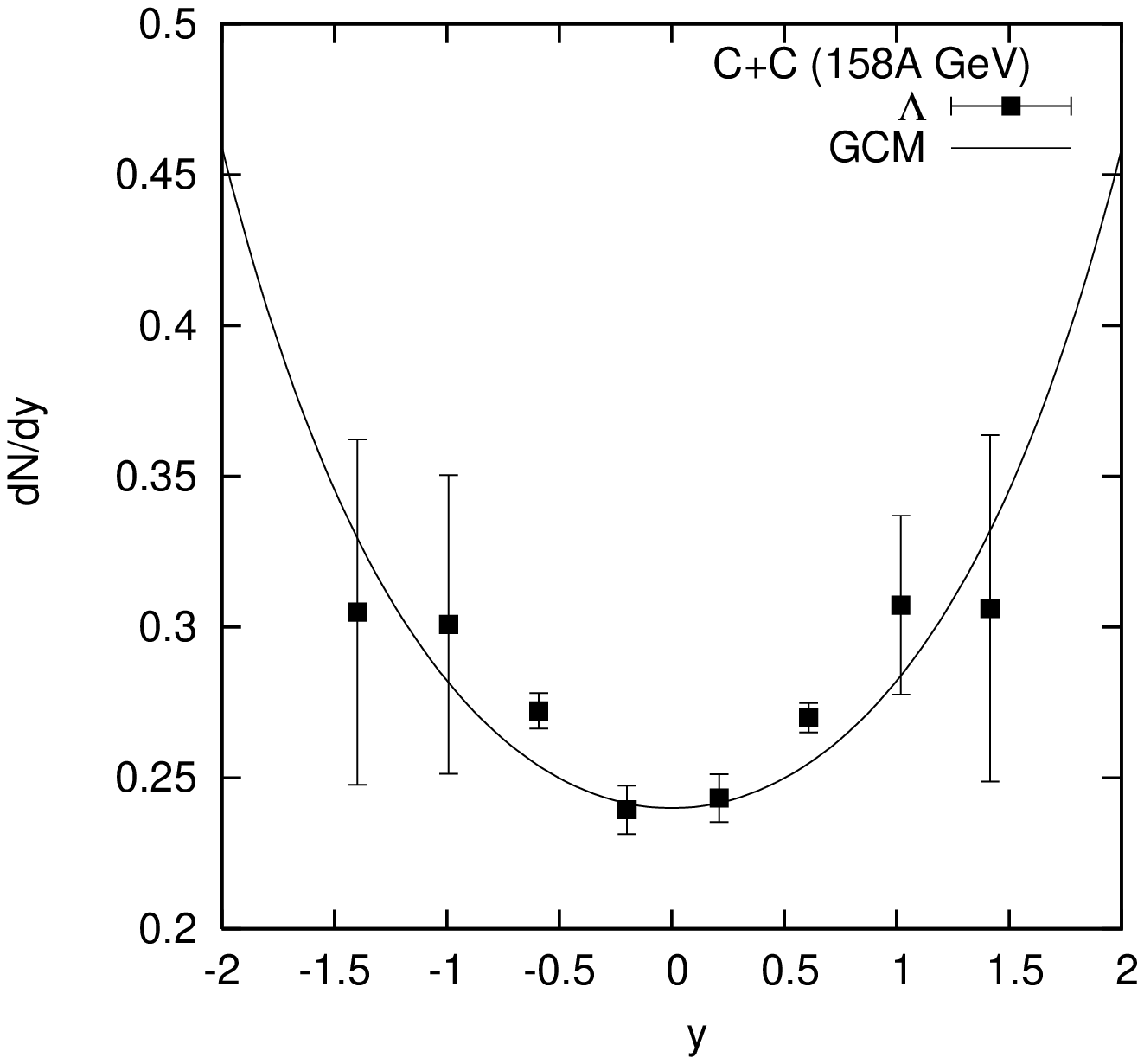}
\end{minipage}}%
\subfigure[]{
\begin{minipage}{.5\textwidth}
\centering
 \includegraphics[width=2.5in]{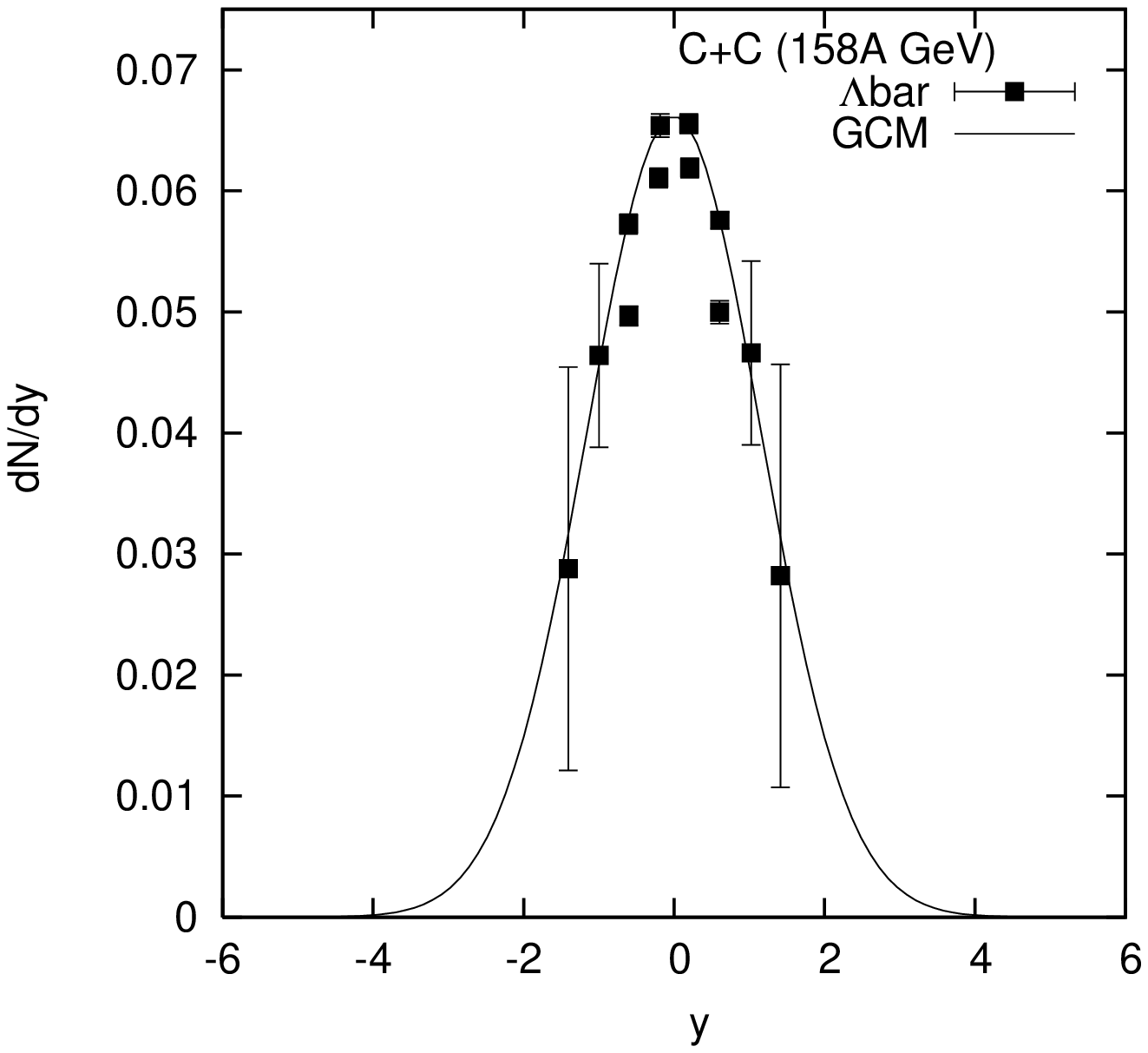}
 \end{minipage}}%
\caption{The rapidity spectra of $\Lambda$ and $\Lambda$bar for near central C+C collisions at 158A GeV
for $\beta$=0. The experimental points are taken from \cite{Anticic1} and the parameter
values are taken from Table 10. The solid curve provide the GCM-based results.}
%\end{figure}

%\begin{figure}
\subfigure[]{
\begin{minipage}{.5\textwidth}
\centering
\includegraphics[width=2.5in]{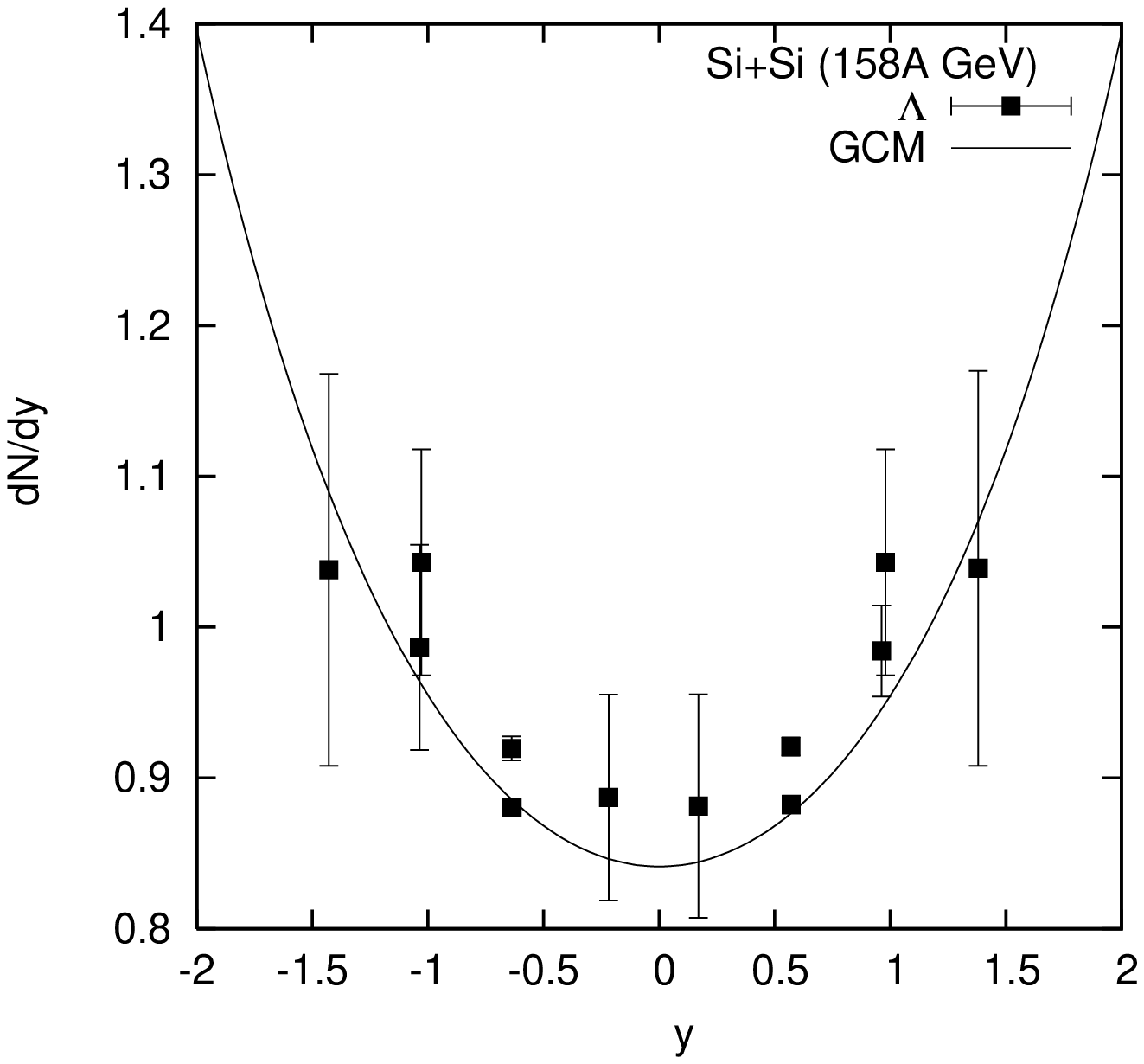}
\end{minipage}}%
\subfigure[]{
\begin{minipage}{.5\textwidth}
\centering
 \includegraphics[width=2.5in]{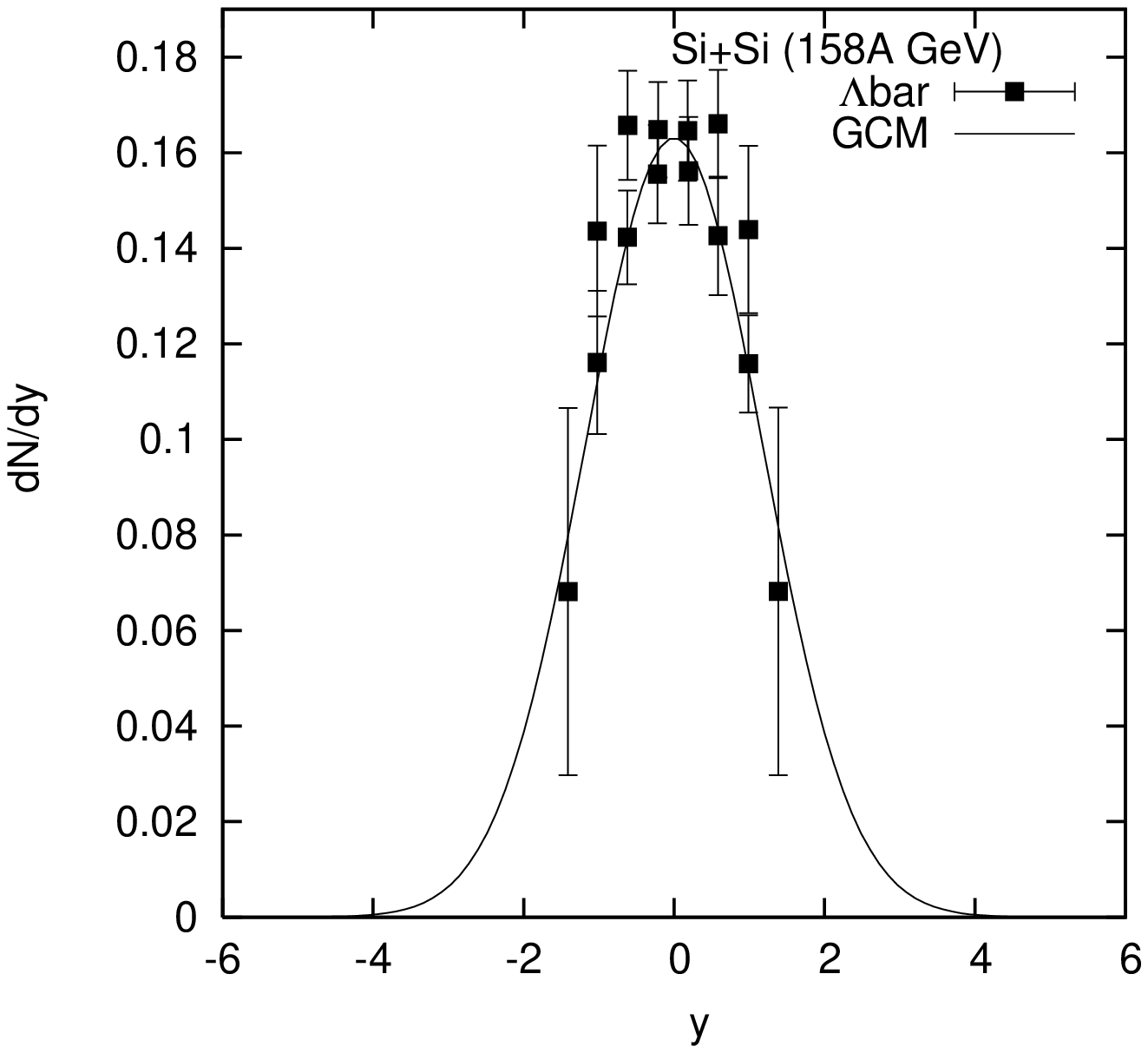}
 \end{minipage}}%
\caption{The rapidity spectra of $\Lambda$ and $\Lambda$bar for near
central Si+Si collisions at 158A GeV for $\beta$=0. The experimental
points are taken from \cite{Anticic1} and the parameter values are
taken from Table 10. The solid curve provide the GCM-based results.}
\end{figure}

\end{document}